\documentclass[aps,pra,twocolumn,superscriptaddress, biblatex]{revtex4-2}

\pdfoutput=1
\usepackage[english]{babel}

\usepackage[colorlinks=true, breaklinks=true]{hyperref}
\usepackage{epsfig,graphicx}
\usepackage{indentfirst}
\usepackage{amsfonts,amssymb,latexsym,amsmath,enumerate,amsthm}
\usepackage{bm}
\usepackage{color}
\usepackage{appendix}

\usepackage{dcolumn}

\usepackage{url}

\usepackage{physics}
\usepackage{soul}

\usepackage[sort&compress]{natbib}
\begin{document}

% \title{The use of electron scattering by atoms to analyze  \\  relativistic twisted electron beams}
% Study of  relativistic twisted electrons by atomic scattering
\title{% Analyzing the relativistic twisted electrons by an atomic scattering\\
 %Study of highly relativistic vortex electron beams by atomic scattering
 Studying highly relativistic vortex-electron beams by atomic scattering}

\author{V.\,K.~Ivanov}
    \email[E-mail: ]{vladislav.ivanov@metalab.ifmo.ru}
\affiliation{School of Physics and Engineering, ITMO University, 
197101 St.\,Petersburg, Russia}
% \affiliation{School of Physics and Engineering, Faculty of Physics, ITMO University, Kronverksky Pr. 49, bldg. A, St. Petersburg, 197101, Russia}

\author{ A.\,D.~Chaikovskaia}
\affiliation{School of Physics and Engineering, ITMO University, 
197101 St.\,Petersburg, Russia}

\author{D.\,V.~Karlovets}
    % \email[E-mail: ]{d.karlovets@gmail.com}
\affiliation{School of Physics and Engineering, ITMO University, 
197101 St.\,Petersburg, Russia}

\date{\today}

\begin{abstract}
We explore the opportunities of using electron scattering by screened Coulomb potential as a tool to retrieve properties of the relativistic vortex beams of electrons, such as their transverse momentum and orbital angular momentum (OAM). We focus on relativistic and ultra-relativistic regimes of the electron energies of at least several MeV and higher, in which the transverse beam momentum is typically much smaller than its longitudinal momentum. Different scattering scenarios for the incident electron beam are considered. In particular, the scattering by a very wide target can be used to probe the electron transverse momentum when its values are larger than 10 keV. The scattering by a target of a width comparable to that of the incident beam allows one to obtain information about the electron OAM. Varying target sizes in the range from couple to hundreds of nanometers, one can in principle distinguish OAM values from several units of $\hbar$ up to thousands and more.
 \end{abstract}

\maketitle

\section{Introduction} 

Particles with a definite value of orbital angular momentum (OAM) \cite{torres2011twisted, Ivanov2011, BLIOKH20171}, also dubbed \textit{twisted} or \textit{vortex}  particles, are of considerable attention nowadays. Historically, twisted photons were first to be studied both in  theoretical and experimental domains, and pioneering experiments with them took place in the 1990s. Generation of twisted electrons is a more recent research field --  first such electrons were obtained in the early 2010s \cite{uchida2010, verbeeck2010, McMorran2011}. Currently, the achievable values of OAM projection value could be as high as hundreds \cite{Grillo2015} and even thousands \cite{mafakheri, McMorran2017} 
of units in terms of $\hbar$. Applications of twisted electrons include such spheres as ionization by twisted electrons \cite{Plumadore_2020, Harris_2019, Dhankhar_2020} and interaction of twisted electrons with matter \cite{Lloyd2012, LLoyd2012_Dichroism}. 

Keeping up with the experiment, quantitative theoretical studies of the scattering processes with twisted electrons are in development. Let us briefly review several works.
% Naturally, scattering of plane wave electrons by atomic potentials is a textbook material. 
Scattering of twisted electrons by single potential atomic targets and  infinitely wide (\textit{macroscopic})  targets were considered in \cite{serbo2015, Kosheleva2018} in non-relativistic and moderately relativistic regimes (electron kinetic energy up to $1$~MeV). A more sophisticated approach to twisted states consists in treating them as spatially localized wave–packets. Scattering of an  “ordinary” Gaussian packet by a single atom, macroscopic or a localized finite size (\textit{mesoscopic}) targets is given in \cite{karlovets2015}.
Generalizations for the case of twisted particles in non-relativistic regime could be found in \cite{karlovets2017, VanBoxem2014, VanBoxem2015}. Here we shall consider the scattering processes with relativistic energies for a single atom, macroscopic and mesoscopic targets.

% 100 nm target atom, ring size should be taken into account, overlap

In this work, we revisit the topic of using electron scattering by an atomic potential as a tool for analyzing properties of the relativistic Bessel beams of electrons.
It is argued in \cite{serbo2015} that the case of scattering by single atom is rather informative on the features of the incident twisted particle, and it is even possible to retrieve the value of the OAM projection, while, in contrast, scattering by macroscopic target is less sensitive: transverse momentum could be deduced but not the OAM value.
In the same time, calculations with a single atom target are not straightforwardly applicable  in a real experiment. 

%To overcome this problem, 
To have both realistic and OAM sensitive scenario, we suggest taking a finite size target, the \textit{mesoscopic} target, following the example of \cite{karlovets2017}.
%for scattering of a relativistic twisted electron. 
We find an amplitude for the scattering off a mesoscopic target and compare it to the one off a macroscopic target. 
As the mesoscopic target  continues naturally both into the single atom and macroscopic scenarios, it provides a signature criterion to mark the transition between these two scenarios. We find that the transition takes place at different target sizes depending on the OAM of the incident twisted electron, and develop an idea that this can be used to retrieve the value of  OAM in experiment.
%The obtained expression shows that the transition “point” from the single atom case to the macroscopic one takes place at different target sizes depending on the OAM of the twisted electron.} 
%We develop an idea that this can be used to retrieve the value of twisted electron OAM in experiment.
Having in mind the possibilities of generating relativistic twisted electrons at particle accelerators \cite{Karlovets_2021_NJP, Karlovets2022_gen, IVANOV2022103987}, we pay specific attention to the ultra-relativistic energies, starting from several MeV and higher. 
For such energies the usual methods of analyzing the twisted electron beams used in electron microscopy, in which the typical electron energies are of order of several keVs, are hardly applicable, and that makes the proposed method of detecting OAM promising for analyzing relativistic electron beams. 

In Section \ref{section_ii} we review the basics of Mott scattering, the modifications needed for the study of twisted electron scattering and the technical realizations of the three aforementioned target kinds.
After these introductory steps, we find scattering amplitudes for all three scenarios in Section \ref{section_iii}. The results acquired are analyzed in Section \ref{section_iv} and, finally, the summary is presented in Section \ref{section_v}.

Through the paper we put $\hbar = c =1$ and use  Gaussian convention for the electric charge: $\alpha_0 = e^2 = 1/137$.
%unrationalized
% \section{Theoretical formalism}\label{section_ii}

\section{Theoretical preliminaries}\label{section_ii}

\subsection{Plane-wave Mott scattering}

The Mott scattering description is given in many classical textbooks \cite{akhiezer1965quantum, berestetskii}. The corresponding scattering amplitude can be written as:

\begin{eqnarray}
    &f_{\lambda,\lambda^\prime}(\vb{p},\vb{p}^\prime)=-\int \psi_{\vb{p}^\prime \lambda^\prime}^\dagger(\vb{r})\mathcal{V}(\vb{r})\psi_{\vb{p} \lambda}(\vb{r}) \dd^3r,
    \label{plane_amp_def}\\
    &S_{fi} = i 2\pi \delta(\varepsilon - \varepsilon^\prime) f_{\lambda,\lambda^\prime}(\vb{p},\vb{p}^\prime),
\end{eqnarray}

\noindent where 

\begin{equation}
    \psi_{\vb{p} \lambda} = \frac{1}{\sqrt{2\varepsilon V}}u_{\vb{p} \lambda}e^{i\vb{p}\vb{r}},\,\psi_{\vb{p}^\prime \lambda^\prime} = \frac{1}{\sqrt{2\varepsilon^\prime V}} u_{\vb{p}^\prime \lambda^\prime}e^{i\vb{p}^\prime\vb{r}}
    \label{plane_def}
\end{equation}

\noindent are the plane-wave  wave-functions of free electrons with incident (final) momentum, energy and helicity $\vb{p}$, $\varepsilon = \sqrt{\vb{p}^2 + m_e^2}$ and $\lambda$ ($\vb{p}^\prime$, $\varepsilon^\prime$ and $\lambda^\prime$) and $\mathcal{V}(\vb{r})$ is the scattering potential, $m_e$ is the electron mass. The Dirac bispinors $u_{\vb{p}\lambda}$ can be expressed as

\begin{equation}
    u_{\vb{p} \lambda} = \left( \begin{array}{cc}
        \sqrt{\varepsilon+m_e} w^{\lambda}(\vb{n})   \\
        2\lambda\sqrt{\varepsilon-m_e} w^{\lambda}(\vb{n})
    \end{array} \right),
\end{equation}

\noindent where the spinors $w^{\lambda}(\vb{n})$ are the eigenfunctions of the helicity operator and $\vb{n} = (\sin{\theta}\cos{\varphi},\sin{\theta}\sin{\varphi},\cos{\theta})$ is a unit vector along $\vb{p}$: 

\begin{equation}
    \Lambda(\vb{n})w^{\lambda}(\vb{n}) \equiv \frac{\hat{\vb{\sigma}}\vb{n}}{2}w^{\lambda}(\vb{n})=\lambda w^{\lambda}(\vb{n}).
\end{equation}

Let us choice the axes so that the incident electron propagates along $z$ direction. For a spinor $w^{\lambda}$ along the $z$ direction, the relation above becomes

\begin{equation}
    \frac{\hat{\sigma}_z}{2}w^{\sigma}(\vb{e}_z) = \sigma w^{\sigma}(\vb{e}_z).
\end{equation}
In this case, this spinor has simple form for up and down spin:
\begin{equation}
    w^{1/2}(\vb{e}_z) = \left( \begin{array}{cc}
        1 \\
        0
    \end{array}  \right) , \ 
    w^{-1/2}(\vb{e}_z) = \left( \begin{array}{cc}
         0  \\
         1 
    \end{array}  \right)
\end{equation}

There is a standard approach that simplifies further calculation of the twisted particle amplitudes -- representing electron spinors using the Wigner D-functions $D_{\sigma\lambda}^{1/2}(\varphi,\theta,0)$ \cite{varshalovich}: 

\begin{equation}
    \begin{aligned}
    w^{\lambda}(\vb{n}) &= \sum\limits_{\sigma=\pm 1/2} D_{\sigma\lambda}^{1/2}(\varphi,\theta,0)w^{\sigma}(\vb{e}_z) \\
    &= \sum\limits_{\sigma=\pm 1/2} e^{-i\sigma\varphi}d_{\sigma\lambda}^{1/2}(\theta)w^{\sigma}(\vb{e}_z),
    \end{aligned}
\end{equation}

\noindent where $d_{\sigma\lambda}^{1/2}(\theta) = \delta_{\sigma,\lambda}\cos{(\theta/2)} - 2\sigma\delta_{\sigma,-\lambda}\sin{(\theta/2)}$.
The bispinor $u_{\vb{p}\lambda}$ of the incident electron can then be expressed in the following way \cite{Karlovets2022_gen}:

\begin{equation}
    u_{\vb{p}\lambda} = \sum\limits_{\sigma=\pm 1/2} e^{-i\sigma\varphi}d_{\sigma\lambda}^{1/2}(\theta)u_{p_z \sigma}.
\end{equation}

Turning to $\mathcal{V}(\vb{r})$, the Coulomb potential is used for the conventional Mott scattering, but a more accurate result can be obtained using a screened Coulomb potential\footnote{Sometimes this kind of potential is called the Yukawa potential because it has the same functional form.}:

\begin{equation}
    \mathcal{V}(\vb{r})=-\frac{Ze^2}{r}e^{-\mu r},
    \label{potential}
\end{equation}

\noindent where $Z$ is a charge of the nucleus, $e$ is an electron charge and $\mu$ is a parameter of screening, which is set to be equal to $2m_e\alpha_0 = 2/a_0$, where $a_0$ is the Bohr radius, in the case of hydrogen \cite{serbo2015}. After integrating (\ref{plane_amp_def}) with the potential (\ref{potential}) we find

\begin{equation}
    \begin{aligned}
    f_{\lambda,\lambda^\prime}(\vb{p},\vb{p}^\prime) &= 4\pi\frac{Ze^2}{\vb{q}^2+\mu^2}u_{\vb{p}^\prime \lambda^\prime}^\dagger u_{\vb{p}\lambda} \\
    &= 4\pi\frac{2Ze^2}{\vb{q}^2+\mu^2}(\varepsilon\delta_{\lambda\lambda^\prime} + m_e\delta_{\lambda,-\lambda^\prime}) \\
    &\cross \sum\limits_{\sigma=\pm 1/2} e^{i\sigma(\varphi^\prime-\varphi)}d_{\sigma\lambda}^{1/2}(\theta)d_{\sigma\lambda^\prime}^{1/2}(\theta^\prime),
    \end{aligned}
    \label{plane_amp}
\end{equation}

\noindent where primed angles are of the final electron momentum. Here, for elastic scattering we have ($\Theta$ is an angle between momentum vectors):

\begin{equation}
    \begin{aligned}
    \vb{q}^2 &= (\vb{p}-\vb{p}^\prime)^2 = 2\abs{\vb{p}}^2 (1 - \cos{\Theta}) \\
    &= 2\abs{\vb{p}}^2 (1 - \cos{\theta}\cos{\theta^\prime} - \sin{\theta}\sin{\theta^\prime}\cos{(\varphi-\varphi^\prime)})
    \end{aligned}
\end{equation}

In accordance with \cite{berestetskii, serbo2015} the resulting cross-section is
\begin{equation}
    \frac{\dd \sigma}{\dd \Omega} = \frac{\abs{\vb{p}}}{\varepsilon j_{in}}\frac{1}{16\pi^2}\abs{f_{\lambda,\lambda^\prime}(\vb{p},\vb{p}^\prime)}^2,
    \label{cs_def}
\end{equation}

\noindent where $j_{in}$ is a projection on the propagation direction of the incident particle current

\begin{equation}
    j^\mu = \bar{\psi} \gamma^\mu \psi,
    \label{current_def_gen}
\end{equation}

\noindent and $\gamma^\mu$ are Dirac matrices. With the plane wave expression (\ref{plane_def}) for the incident electron substituted into the definition (\ref{current_def_gen}), the identity $\bar{u}_{\vb{p}\lambda}\gamma^\mu u_{\vb{p}\lambda} = 2p^\mu$, and the incident wave propagating along $z$ the z-projection of the current is simplified to:  $j_z = \frac{1}{2\varepsilon} 2p_z = \frac{\abs{\vb{p}}}{\varepsilon}$.
Then, for the cross-section of the plane wave scattering, we find
\begin{equation}
    \begin{aligned}
    \left(\frac{\dd \sigma}{\dd \Omega}\right)^{(PW)} &= \frac{1}{16\pi^2}\abs{f_{\lambda,\lambda^\prime}(\vb{p},\vb{p}^\prime)}^2 \\
    &= \frac{4Z^2e^4}{(\vb{q}^2+\mu^2)^2}(\varepsilon^2\delta_{\lambda\lambda^\prime} + m_e^2\delta_{\lambda,-\lambda^\prime}) \\
    &\cross \abs{\sum\limits_{\sigma=\pm 1/2} e^{i\sigma(\varphi^\prime-\varphi)}d_{\sigma\lambda}^{1/2}(\theta)d_{\sigma\lambda^\prime}^{1/2}(\theta^\prime)}^2.
    \label{pw_cs}
    \end{aligned}
\end{equation}
This cross-section would be useful as a reference in the following discussion.
%We shall not go deep into the discussion of plane wave Mott scattering here, its cross-section is presented here for references that will be useful in our following discussion.

\subsection{Bessel twisted electrons}

%If we restrict the wave function to be eigenfunction of the total angular momentum operator $J_z$ with the eigenvalues $m=0,\pm1,\pm2,...$, then for the particle propagating along $z$ we would find that this wave function will be a “twisted” wave, that is itself a function of Bessel functions of the first kind. This wave function can be written in the following way \cite{serbo2015}:

A twisted electron moving along the $z$ direction is characterized by the value of the total angular momentum (TAM) operator $J_z$, i.\,e. it has a defined value of $m=0,\pm1,\pm2,...$. The corresponding wave function, the so-called {\it Bessel beam}, can be written in the following way \cite{serbo2015}:

\begin{equation}
    \begin{aligned}
    \psi_{\kappa m p_z \lambda}(\vb{r}) &= \int \frac{\dd^2 p_\perp}{(2\pi)^2}a_{\kappa m}(\vb{p}_\perp)\psi_{\vb{p} \lambda} \\
    = \sqrt{\frac{\kappa}{2\pi}} &\sum\limits_{\sigma=\pm 1/2} d_{\sigma\lambda}^{1/2}(\theta_p)u_{p_z \sigma}e^{ip_z z}J_{m-\sigma}(\kappa r) ,
    \end{aligned}
    \label{twisted_def}
\end{equation}
\noindent where $\vb{p}_\perp = (\abs{\vb{p}_\perp}\cos{\varphi_p},\abs{\vb{p}_\perp}\sin{\varphi_p})$ is the transverse part of the electron momentum $\vb{p}$ and $(\vb{p}_\perp, p_z)$ lay on the surface of a cone with an opening angle is $\theta_p=\arctan(\kappa/p_z)$, $J_{m-\sigma}(\kappa r)$ is the Bessel function of the first kind and

\begin{equation}
    a_{\kappa m}(\vb{p}_\perp) = (-i)^m e^{im\varphi_p}\sqrt{\frac{2\pi}{\kappa}}\delta(\abs{\vb{p}_\perp}-\kappa)
\end{equation}

\noindent is a Fourier coefficient with $\kappa$ fixing the modulus of the transverse momentum. In the limit of $\kappa \rightarrow 0 $ the twisted wave function (\ref{twisted_def}) behaves like a plane wave $\psi_{p_z \lambda}$.

It is worth noting, that a more general  model for a twisted electron is the Laguerre-Gaussian (LG) beam that takes into account the beam spreading. However, since Rayleigh length of electron LG beam %\pi w_0^2/\lambda$, where $w_0$ is the waist radius and $\lambda$ is the wavelength, 
is much larger than the typical scale in our settings, the beam spreading is actually negligible. Moreover, in experiment only several first rings of the LG beam profile contribute, therefore the Bessel beam is a good approximation of the twisted beam profile.

The amplitude for the scattering of the initial twisted electron is obtained via inserting the twisted wave function (\ref{twisted_def}) into Eq.~(\ref{plane_amp_def}):
\begin{equation}
    \begin{aligned}
    f_{\lambda,\lambda^\prime}^{m,(TW)}(\vb{p},\vb{p}^\prime,\vb{b}) = -\int \psi_{\vb{p}^\prime \lambda^\prime}^\dagger(\vb{r})\mathcal{V}(\vb{r})&\psi_{\kappa m p_z \lambda}(\vb{r}) \dd^3r \\
    = \int \frac{\dd^2 p_\perp}{(2\pi)^2}a_{\kappa m}(\vb{p}_\perp)e^{-i\vb{p}_\perp\vb{b}}&f_{\lambda\lambda^\prime}(\vb{p},\vb{p}^\prime) \\
    = (-i)^m \sqrt{\frac{2\pi}{\kappa}} \int\limits_0^{2\pi} \frac{\dd \varphi_p}{2\pi}e^{im\varphi_p-i\vb{p}_\perp\vb{b}}&f_{\lambda\lambda^\prime}(\vb{p},\vb{p}^\prime)
    \end{aligned}
    % \label{twisted_amp}
\end{equation}

With $f_{\lambda\lambda^\prime}(\vb{p},\vb{p}^\prime)$ given in Eq.~(\ref{plane_amp}) we obtain

\begin{equation}
    \begin{aligned}
    f_{\lambda,\lambda^\prime}^{m,(TW)}(\vb{p},\vb{p}^\prime,\vb{b})  &=8\pi Ze^2 i^{-m}\sqrt{\frac{\kappa}{2\pi}}(\varepsilon\delta_{\lambda\lambda^\prime} + m_e\delta_{\lambda,-\lambda^\prime})e^{im\varphi^\prime} \\
    \cross &\sum\limits_{\sigma=\pm 1/2} d_{\sigma\lambda}^{1/2}(\theta_p)d_{\sigma\lambda^\prime}^{1/2}(\theta^\prime)\mathcal{I}_{m-\sigma}(\alpha,\beta,\vb{b}),
    \end{aligned}
    \label{twisted_amp}
\end{equation}

\noindent where

\begin{equation}
    \mathcal{I}_{n}(\alpha,\beta,\vb{b}) = \int \frac{\dd\phi}{2\pi}\frac{e^{in\phi-i\kappa b\cos{(\phi+\varphi^\prime-\varphi_b)}}}{\alpha-\beta\cos{\phi}},
    \label{int_def}
\end{equation}

\begin{align}
    \alpha &= 2\abs{\vb{p}}^2 (1-\cos{\theta_p}\cos{\theta^\prime}) + \mu^2,\label{alpha_def}
    \\
    \beta &= 2\abs{\vb{p}}^2 \sin{\theta_p}\sin{\theta^\prime}.
    \label{beta_def}
\end{align}

It seems that the integral (\ref{int_def}) could not be calculated in closed form. However, with some complex analysis and for the case of impact parameter $\vb{b} = 0$  we obtain the following analytical expression:

\begin{equation}
    \begin{aligned}
    \mathcal{I}_{n}(\alpha,\beta,0) &= \int \frac{\dd\phi}{2\pi}\frac{e^{in\phi}}{\alpha-\beta\cos{\phi}} \\
    &= - \frac{1}{i\pi\beta}\oint\limits_{\abs{z}\leq 1} \frac{z^\abs{n}}{(z-\zeta_1)(z-\zeta_2)} \\
    &=  -\frac{2}{\beta}\frac{\zeta_2^\abs{n}}{\zeta_2 - \zeta_1},
    \end{aligned}
    \label{int_b0}
\end{equation}

\noindent where $\zeta_1 >1$, $\zeta_2 < 1$

\begin{equation}
    \zeta_{1,2} = \frac{\alpha}{\beta}\left(1 \pm \sqrt{1 - \frac{\beta^2}{\alpha^2}} \right).
\end{equation}
Eq.~(\ref{int_b0}) can be now expressed as
\begin{equation}
    \mathcal{I}_{n}(\alpha,\beta,0) = \frac{1}{\sqrt{\alpha^2 - \beta^2}} \left( \frac{\alpha - \sqrt{\alpha^2 - \beta^2}}{\beta} \right)^\abs{n}.
\end{equation}

\noindent 
%Notice that the resonant factor $1/ \sqrt{\alpha^2 - \beta^2}$ appears. 
Similar results are given in \cite{VanBoxem2014, serbo2015}.

For small $\theta_p$, the integral (\ref{int_def}) transforms into:

\begin{equation}
\begin{aligned}
    \mathcal{I}_{n}(\alpha,0,\vb{b}) &= \int \frac{\dd\phi}{2\pi}\frac{e^{in\phi-i\kappa b\cos{(\phi+\varphi^\prime-\varphi_b)}}}{\alpha} \\&= \frac{1}{\alpha}e^{-in(\varphi^\prime-\varphi_b)}J_n(\kappa b).
\end{aligned}
    \label{int_small_angle}
\end{equation}

We remark that for the large $\kappa b \gg 1$ the integral (\ref{int_def}) can be evaluated approximately using the method of stationary phase \cite{fedoryuk, bender1999}. We show how it can be done in Appendix \ref{AppB}.

\subsection{Distribution of atoms in the target}

As was mentioned in the introduction we shall analyze electron scattering by the targets of three types - single atom, macroscopic and mesoscopic. Physically, such targets can be represented by a thin foil much wider than the beam size for the macroscopic or a round piece of thin foil for the mesoscopic target. To model a target we average the amplitude (\ref{twisted_amp}) with some distribution function $n(\vb{b})$:

\begin{equation}
    F_{m,\lambda,\lambda^\prime}(\vb{p},\vb{p}^\prime,\vb{b}) = \int \dd^2 \vb{b} \, n(\vb{b})f_{\lambda,\lambda^\prime}^{m,(TW)}(\vb{p},\vb{p}^\prime,\vb{b}).
\end{equation}

\noindent We use the following functions for each target type: 

\begin{align}
    n^{macro}(\vb{b}) &= \frac{1}{\pi R^2} %\theta(|\vb{b}|-R)
    , \label{distr_macro} \\
    n^{single}(\vb{b}) &= \delta(\vb{b}-\vb{b}_0),
    \label{distr_single} \\
    n^{meso}(\vb{b}) &= \frac{1}{2\pi\sigma_b^2} e^{-\frac{1}{2}\left(\frac{\vb{b}-\vb{b}_0}{\sigma_b}\right)^2},
    \label{distr_meso}
\end{align}

\noindent where $R$ is an arbitrary large radius taken as an integration limit in the macroscopic scenario. In the latter two cases: $\vb{b}_0$ is either a target position or a target center, while $\sigma_b$ gives an effective target size. All three distributions obey the normalization condition $\int \dd^2 b \, n(\vb{b}) = 1$.

The mesoscopic distribution (\ref{distr_meso}) should reduce to (\ref{distr_macro}) for a large target size ($\sigma_b \rightarrow \infty$); we have:
\begin{equation}
    n^{meso}(\vb{b}) \approx \frac{1}{2\pi\sigma_b^2} \sim n^{macro}(\vb{b}). 
\end{equation}
In the limit of small $\sigma_b$, the mesoscopic distribution is reduced to the Dirac delta.

% In Appendix \ref{AppB} we show, how an alternative distribution (the uniform distribution over a finite interval) can be used to model a mesoscopic target.

\section{Scattering off different targets}\label{section_iii}

In this section, we present the results for the scattering of a twisted electron by different targets. Results for the single atom and the macroscopic target are similar to those presented in \cite{serbo2015}. Though, we are more interested in the relativistic energies of the incident electron and focus on the regime when $\theta_p$ is rather small ($p_z\gg\kappa$) or, equivalently, when the factor $\beta$ of Eq.~(\ref{beta_def}) is small. For instance, ultra-relativistic electrons with  $\varepsilon = 100 m_e$ and $\kappa \sim 10$ keV  have the value of angle $\theta_p \approx$ $0.01^\circ$, and therefore $\alpha \gg \beta$. We assume, that the electron scatters on different atoms independently, and obtain scattering amplitude by summing the contributions of scattering by individual potentials. This approach  seems to be adequate for the beam size typically larger than the atomic scale. For the specifics of considering more tightly focused beams we refer the reader to  Ref.~\cite{KarlovetsPRL2017}.

We also provide calculations for the relativistic electron scattering by a mesoscopic target, which are complementary to those made in  \cite{karlovets2017} in the non-relativistic regime. 
%The results for the mesoscopic target were obtained in \cite{karlovets2017} for non-relativistic electron; here, we study the relativistic case.

\subsection{Macroscopic target}

The simplest case is scattering by a macroscopic target. In this scenario, we can obtain an analytic expression for the cross-section. 
Let us start with squaring the corresponding amplitude:
\begin{equation}
    \begin{aligned}
    \abs{F_{m,\lambda,\lambda^\prime}^{(macro)}(\vb{p},\vb{p}^\prime)}^2 &= \\ =
     \frac{1}{\pi R^2}\frac{2\pi}{\kappa}\int &\frac{\dd^2 \vb{p}_\perp}{(2\pi)^2} \abs{f_{\lambda\lambda^\prime}(\vb{p},\vb{p}^\prime)\delta(\abs{\vb{p}_\perp}-\kappa)}^2,
    \end{aligned}
\end{equation}

\noindent where we used that $\int \dd^2 b \, e^{i(\vb{k}_\perp-\vb{p}_\perp)\vb{b}} = (2\pi)^2\delta(\vb{k}_\perp-\vb{p}_\perp)$.
Further, one can use the following identity (see, for example, \cite{Jentschura_2011}):
\begin{equation}
    \abs{\delta(\abs{\vb{p}_\perp}-\kappa)}^2 = \frac{R}{\pi} \delta(\abs{\vb{p}_\perp}-\kappa).
\end{equation}
% \noindent where $R$ is, again, a large radius integration limit.
Then it turns out that
\begin{equation}
    \begin{aligned}
    \abs{F_{m,\lambda,\lambda^\prime}^{(macro)}(\vb{p},\vb{p}^\prime)}^2 
    &= \frac{1}{R}64Z^2e^4 (\varepsilon^2\delta_{\lambda\lambda^\prime} + m_e^2\delta_{\lambda,-\lambda^\prime}) \\
    &\cross \sum\limits_{\sigma,\sigma^\prime} d_{\sigma\lambda}^{1/2}(\theta_p)d_{\sigma\lambda^\prime}^{1/2}(\theta^\prime)d_{\sigma^\prime\lambda}^{1/2}(\theta_p)d_{\sigma^\prime\lambda^\prime}^{1/2}(\theta^\prime) \\
    &\cross \int\limits_0^{2\pi} \frac{\dd \varphi_p}{2\pi} \frac{e^{i(\sigma - \sigma^\prime)(\varphi_p - \varphi^\prime)}}{(\alpha - \beta \cos{(\varphi_p - \varphi^\prime)})^2}.
    \end{aligned}
\end{equation}
The integral in the above expression can be expressed through the integral (\ref{int_b0}):

\begin{equation}
\begin{aligned}
    \int\limits_0^{2\pi} \frac{\dd \varphi_p}{2\pi} \frac{e^{i(\sigma - \sigma^\prime)(\varphi_p - \varphi^\prime)}}{(\alpha - \beta \cos{(\varphi_p - \varphi^\prime)})^2} &= -\frac{\partial}{\partial \alpha} \mathcal{I}_{\sigma^\prime-\sigma}(\alpha,\beta,0) \\
    &= \frac{\alpha\delta_{\sigma,\sigma^\prime} + \beta\delta_{\sigma,\sigma^\prime}}{(\alpha^2-\beta^2)^{3/2}}.
\end{aligned}   
\end{equation}

To find the cross-section we need to calculate the incident electron current $j_z$. In contrast to the plane wave case, for a twisted wave-function we need to average the current over the incident plane \cite{serbo2015}:

\begin{equation}
\begin{aligned}
    &j_z^{(macro)} = \frac{1}{\pi R^2}\int \dd^2 b \, \bar{\psi}_{\kappa m p_z \lambda}(\vb{b}) \gamma^3 \psi_{\kappa m p_z \lambda}(\vb{b}) \\
    &= \frac{1}{2\varepsilon\pi R^2}\frac{\kappa}{2\pi} \int \dd^2 b \, \sum\limits_{\sigma,\sigma^\prime}d_{\sigma\lambda}^{1/2}(\theta_p)d_{\sigma^\prime\lambda}^{1/2}(\theta_p)\bar{u}_{p_z \sigma} \gamma^3 u_{p_z \sigma^\prime} \\
    &\cross \int\limits_0^{2\pi}\frac{\dd \varphi_p}{2\pi} e^{i(m-\sigma^\prime)\varphi_p - i\kappa b \cos{\varphi_p}}\int\limits_0^{2\pi}\frac{\dd \varphi_p}{2\pi} e^{-i(m-\sigma)\varphi_p + i\kappa b \cos{\varphi_p}}  \\
    &= \frac{1}{\pi R^2}\frac{p_z}{\varepsilon} \int_0^\infty \dd b \, \kappa b J_{m-\lambda}^2(\kappa b) = \frac{p_z}{\varepsilon} \frac{1}{\pi^2R}
\end{aligned}
\label{current_macro}
\end{equation}

\begin{figure*}[!htb]
% \begin{figure*}[h]
    \centering
    \includegraphics[width=\linewidth]{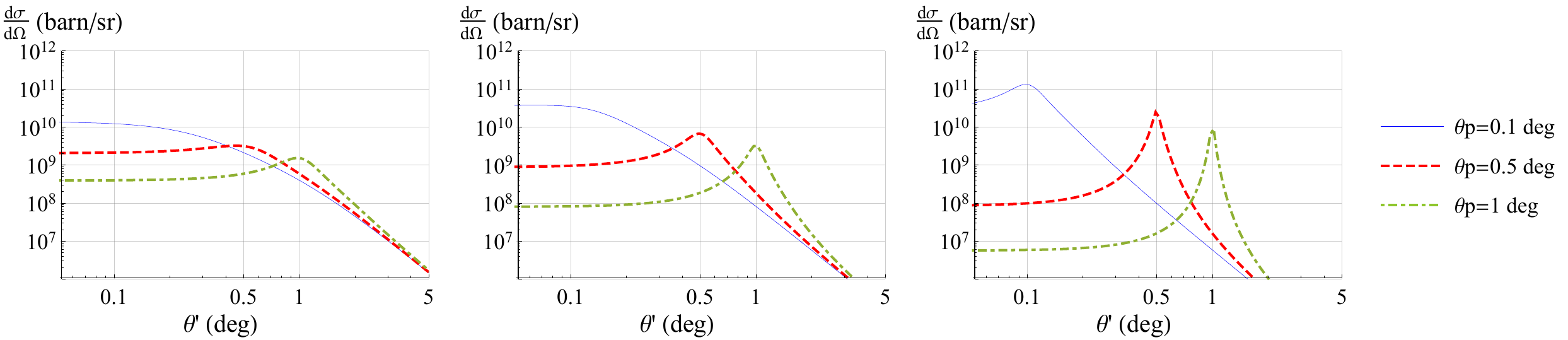}
    \caption{The cross-section for a macrosopic iron target and relativistic electron energies. \textit{Left panel}: $\varepsilon = 2m_e$, blue solid $\theta_p = 0.1^\circ$ ($\kappa = 1.5$ keV), dashed red   $\theta_p = 0.5^\circ$  ($\kappa = 7.7$ keV), dot dashed green  $\theta_p = 1^\circ$  ($\kappa = 15$ keV); \textit{Middle panel}: $\varepsilon = 5m_e$, blue solid   $\theta_p = 0.1^\circ$  ($\kappa = 4.4$ keV), dashed red   $\theta_p = 0.5^\circ$  ($\kappa = 21.9$ keV), dot dashed green  $\theta_p = 1^\circ$  ($\kappa = 43$ keV); \textit{Right panel}: $\varepsilon = 20 m_e$, blue solid   $\theta_p = 0.1^\circ$  ($\kappa = 17.8$ keV), dashed red   $\theta_p = 0.5^\circ$  ($\kappa = 89$ keV), dot dashed green  $\theta_p = 1^\circ$  ($\kappa = 178$ keV).
    }
    \label{fig:macro_fe}
\end{figure*}

\begin{figure}[!htb]
    \centering
    \includegraphics[width=0.7\linewidth]{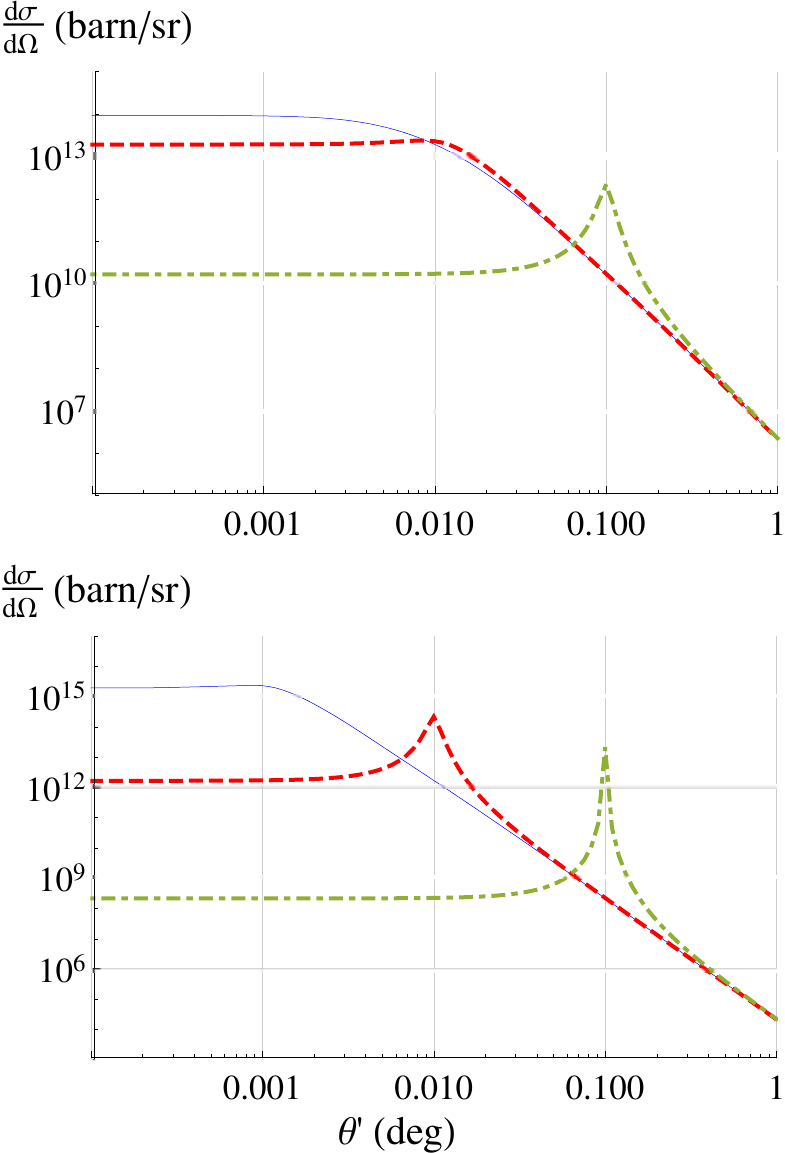}
    \caption{The cross-section for a macrosopic golden target and ultra-relativistic electron energies. \textit{Top panel}: $\varepsilon = 100m_e$, blue solid $\theta_p = 0.001^\circ$ ($\kappa = 0.89$ keV), dashed red   $\theta_p = 0.01^\circ$  ($\kappa = 8.9$ keV), dot dashed green  $\theta_p = 0.1^\circ$  ($\kappa = 89$ keV); \textit{Bottom panel}: $\varepsilon = 1000m_e$, blue solid   $\theta_p = 0.001^\circ$  ($\kappa = 8.9$ keV), dashed red   $\theta_p = 0.01^\circ$  ($\kappa = 89$ keV), dot dashed green  $\theta_p = 0.1^\circ$  ($\kappa = 891$ keV).
    }
    \label{fig:macro_au_he}
\end{figure}

Using Eq.~(\ref{cs_def}) for the cross-section, we find

\begin{equation}
    \begin{aligned}
    \left(\frac{\dd \sigma}{\dd \Omega}\right)^{(macro)} &= \frac{4Z^2e^4}{\cos{\theta_p}} (\varepsilon^2\delta_{\lambda\lambda^\prime} + m_e^2\delta_{\lambda,-\lambda^\prime}) \\
    &\cross \sum\limits_{\sigma,\sigma^\prime} d_{\sigma\lambda}^{1/2}(\theta)d_{\sigma\lambda^\prime}^{1/2}(\theta^\prime)d_{\sigma^\prime\lambda}^{1/2}(\theta)d_{\sigma^\prime\lambda^\prime}^{1/2}(\theta^\prime) \\
    &\cross \frac{\alpha\delta_{\sigma,\sigma^\prime} + \beta\delta_{\sigma,-\sigma^\prime}}{(\alpha^2-\beta^2)^{3/2}} .
    \end{aligned}
    \label{mac_cs}
\end{equation}
Notice the factor $\sim 1/(\alpha^2-\beta^2)$ in the expression above. It leads to the appearance of a characteristic resonance peak in the cross-section graph (see Figures \ref{fig:macro_fe} and \ref{fig:macro_au_he}) , its position around $\theta'\simeq \theta_p$.

%In the relativistic regime, the angle between the incident momentum vector $(\vb{p}_\perp, p_z)$ and the $z$-axis (or momentum cone angle) $\theta_p$ is typically small.  
Let us rewrite the plane wave expression (\ref{pw_cs}) in different terms for convenience:
\begin{equation}
    \begin{aligned}
    \left(\frac{\dd \sigma}{\dd \Omega}\right)^{(PW)}
    &= \frac{4Z^2e^4}{\alpha^2}(\varepsilon^2\delta_{\lambda\lambda^\prime} + m_e^2\delta_{\lambda,-\lambda^\prime}) \\
    &\cross \abs{\sum\limits_{\sigma=\pm 1/2} e^{i\sigma(\varphi^\prime-\varphi)}d_{\sigma\lambda}^{1/2}(\theta)d_{\sigma\lambda^\prime}^{1/2}(\theta^\prime)}^2
    \end{aligned}
    \label{pw_cs2}
\end{equation}
Then, if we go into the relativistic regime in (\ref{mac_cs}) --   neglecting the $\beta$ terms  and assuming $\cos{\theta_p} = 1$ -- we find coincidence with (\ref{pw_cs2}):

\begin{equation}
    \begin{aligned}
    \lim\limits_{\theta_p \rightarrow 0}&\left(\frac{\dd \sigma}{\dd \Omega}\right)^{(macro)} = 4Z^2e^4 (\varepsilon^2\delta_{\lambda\lambda^\prime} + m_e^2\delta_{\lambda,-\lambda^\prime})\cross \\
    &\cross \abs{\sum\limits_{\sigma=\pm 1/2} d_{\sigma\lambda}^{1/2}(\theta_p)d_{\sigma\lambda^\prime}^{1/2}(\theta^\prime)}^2 \frac{1}{\alpha^2} = \left(\frac{\dd \sigma}{\dd \Omega}\right)^{(PW)}.
    \end{aligned}
    % \label{mac_cs}
\end{equation}
This expression is  approximately valid for the values of $\theta_p \lesssim 5$~deg, but near the resonance the Taylor series converges much slower. 
%Therefore, near this resonance the small angle approximation does not work well, so we can expect a difference between the approximation and the exact solution for the twisted particle cross-section, even for a small angle near the resonance. 
On the other hand, for small $\kappa \lesssim 10$ keV the resonance is barely distinguishable (see Figure \ref{fig:macro_fe}).
Hence, the macroscopic cross-section is well-approximated by the plane wave one for $\theta_p \lesssim 5$~deg  when we are not interested in scattering angles near the resonance or when $\kappa$ is small.
%Hence, this problem can be ignored for a small $\kappa$ or when we are not interested in scattering angles near the resonance.

In the discussion above, we used the screened Coulomb potential (\ref{potential}). However, as was proposed in \cite{serbo2015}, we can use an alternative form of the potential -- an analytical fit to the self-consistent Dirac--Hartree--Fock--Slater data \cite{salvat1987}. It is assumed to provide a more realistic description of atom-electron collisions. Such a potential would read:
%analytical fitting procedure to Dirac-Hartree-Fock-Slater (DHFS) self-consistent data
\begin{equation}
    \mathcal{V}_{at}(r) = - \frac{Ze^2}{r} \sum\limits_{i=1}^3 A_i e^{-\mu_i r},
    \label{potential_at}
\end{equation}

\noindent where the coefficients $A_i$ and $\mu_i$ depend on the atomic number and are given in \cite{salvat1987}, $A_1 + A_2 + A_3 = 1$.  See the coefficients for iron and gold in Table \ref{tab:coef}. 

\begin{table}[!htb]
\footnotesize
\caption{\label{tab:coef}%
Parameters of atomic potential (\ref{potential_at}) for some elements \cite{serbo2015, salvat1987}.
}
\begin{ruledtabular}
\begin{tabular}{c|c|c|c|c|c}
% \colrule
Elem. & $A_1$ & $A_2$ & $\mu_1/(m_e \alpha_0)$ & $\mu_2/(m_e \alpha_0)$ & $\mu_3/(m_e \alpha_0)$ \\
\hline Fe(26) & 0.0512 & 0.6995 & 31.825 & 3.7716 & 1.1606 \\
Cu(29) & 0.0771 & 0.7951 & 25.326 & 3.3928 & 1.1426
\\
Ag(47) & 0.2562 & 0.6505 & 15.588 & 2.7412 & 1.1408
\\
Au(79) & 0.2289 & 0.6114 & 22.864 & 3.6914 & 1.4886 \\
\end{tabular}
\end{ruledtabular}
\end{table}

We can implement this new potential into our calculations.
For scattering by a macroscopic target, the result for the cross-section was obtained in \cite{serbo2015}. Repeating all the previous steps we find:

\begin{equation}
    \begin{aligned}
    \left(\frac{\dd \sigma}{\dd \Omega}\right)^{(macro)}_{at} &= \frac{4Z^2e^4}{\cos{\theta_p}} (\varepsilon^2\delta_{\lambda\lambda^\prime} + m_e^2\delta_{\lambda,-\lambda^\prime}) \\
    &\cross \sum\limits_{\sigma,\sigma^\prime} d_{\sigma\lambda}^{1/2}(\theta)d_{\sigma\lambda^\prime}^{1/2}(\theta^\prime)d_{\sigma^\prime\lambda}^{1/2}(\theta)d_{\sigma^\prime\lambda^\prime}^{1/2}(\theta^\prime) \\
    &\cross \sum\limits_{i,k=1}^3 A_i A_k \mathcal{I}_{\sigma-\sigma^\prime}^{at}(\alpha_i,\alpha_k,\beta),     \label{mac_cs_at}
    \end{aligned}
\end{equation}

\begin{equation}
    \mathcal{I}_{n}^{at}(\alpha_i,\alpha_k,\beta) = 
    \left\lbrace \begin{aligned}
    &\frac{\alpha_i\delta_{\sigma,\sigma^\prime} + \beta\delta_{\sigma,\sigma^\prime}}{(\alpha_i^2-\beta^2)^{3/2}}, \text{\,\,\,\,\,\,\,\,\,\,\,\,\,\,\,\,\,\,\,\,\,\,\,\,\,\, if $i = k$} \\
    &\frac{\mathcal{I}_{n}(\alpha_i,\beta,0)-\mathcal{I}_{n}(\alpha_k,\beta,0)}{\alpha_k - \alpha_i}, \text{ if $i \neq k$}
    \end{aligned}\right.  
\end{equation}

Since $\mathcal{I}_{n}(\alpha,\beta,0)$ is proportional to $1/\sqrt{\alpha^2 - \beta^2}$, so is the function $\mathcal{I}_{\sigma-\sigma^\prime}^{at}(\alpha_i,\alpha_k,\beta)$. Therefore, the cross-section (\ref{mac_cs_at}) with the modified potential manifests the same resonance as Eq.~(\ref{mac_cs}).

For small $\theta_p$, the cross-section (\ref{mac_cs_at}) is reduced to:

\begin{equation}
    \begin{aligned}
    \lim\limits_{\theta_p \rightarrow 0}&\left(\frac{\dd \sigma}{\dd \Omega}\right)^{(macro)}_{at} = 4Z^2e^4 (\varepsilon^2\delta_{\lambda\lambda^\prime} + m_e^2\delta_{\lambda,-\lambda^\prime})\cross \\
    &\cross \left(\sum\limits_{i=1}^3 \frac{A_i}{\alpha_i}\right)^2 \abs{\sum\limits_{\sigma=\pm 1/2} d_{\sigma\lambda}^{1/2}(\theta_p)d_{\sigma\lambda^\prime}^{1/2}(\theta^\prime)}^2 
    \end{aligned}
    % \label{mac_cs}
\end{equation}

Shall we deal with a detector, that is insensitive to the electron polarization, the sum of the cross-section over final polarizations $\lambda^\prime$ is necessary.
%Then for such a detector, we shall detect $\sum_{\lambda^\prime} \dd\sigma/\dd \Omega$.

\subsection{Single atom target}

Now we shall analyze the scattering by a single atom potential.
%This case is more complicated than the macroscopic target scattering studied above because finding an analytic result for the amplitude is not possible (or at least much harder). Furthermore, defining a cross-section here is not as obvious as for the macroscopic target or for the plane wave scattering \cite{serbo2015}.
The distribution function for the scattering by a single atom is given by Eq.~(\ref{distr_single}), and the squared scattering amplitude then reads

\begin{equation}
    \begin{aligned}
    &\left|F_{m,\lambda,\lambda^\prime}^{(single)}(\vb{p},\vb{p}^\prime)\right|^2 
    =  \abs{F(\vb{p},\vb{p}^\prime,\vb{b}_0)}^2 \\
    &= 32\pi Z^2e^4 \kappa(\varepsilon^2\delta_{\lambda\lambda^\prime} + m_e^2\delta_{\lambda,-\lambda^\prime}) \\
    &\cross \abs{\sum\limits_{\sigma=\pm 1/2} d_{\sigma\lambda}^{1/2}(\theta_p)d_{\sigma\lambda^\prime}^{1/2}(\theta^\prime)\mathcal{I}_{m-\sigma}(\alpha,\beta,\vb{b}_0)}^2.  
    \end{aligned}
    \label{single_amp}
\end{equation}

In contrast to the previous cases, obtaining an analytic result here is obstructed due to an ambiguity in defining the cross-section. The problem is that the incident current $j_z$ goes to zero for some values of the impact parameter since $j_z \sim J_{m-\lambda}^2(\kappa b_0)$, and hence the usage of Eq.~(\ref{cs_def}) is compromised (see also a discussion on this issue given in \cite{serbo2015}). To overcome this obstacle, we go instead to the consideration of the number of scattering events, following the lead of \cite{karlovets2017,karlovets2022}:

%We would try to find the cross-section for twisted electron scattering by a single atom, however, in this case, the incident current $j_z$ value goes to zero for some values of the impact parameter, since $j_z \sim J_{m-\lambda}^2(\kappa b_0)$, hence defining the cross-section is questionable for such target \cite{serbo2015}.

%To overcome the difficulties with defining the cross-section for scattering by a single atom and to somehow describe its scattering properties, we can use the number of scattering events instead \cite{karlovets2017,karlovets2022}.

\begin{equation}
    \dd \nu \equiv N_e \abs{S_{fi}}^2 \frac{V\dd^3 p^\prime}{(2\pi)^3},
\end{equation}

\noindent where $N_e$ is a number of incident electrons. Then we acquire a relation similar to the cross-section formula

\begin{equation}
    \left(\frac{\dd \nu}{\dd \Omega}\right)^{(single)} = \frac{N_e}{16\pi^2}\frac{\abs{\vb{p}}}{\varepsilon} \abs{F_{m,\lambda,\lambda^\prime}^{(single)}(\vb{p},\vb{p}^\prime)}^2.
    \label{noe_single_1}
\end{equation}

The \textit{luminosity} \cite{Schwartz2014,karlovets2017}  then reads

\begin{equation} \label{lum_def}
    L^{(TW)} = N_e \frac{\kappa}{2 \pi} \frac{\abs{\vb{p}}}{\varepsilon} \frac{\pi T}{R L_z} = N_e \frac{\kappa}{2 \pi}\frac{\pi \abs{\vb{p}} }{R v \varepsilon} =
    \frac{N_e}{\cos{\theta_p}} \frac{\kappa}{2 \pi}\frac{\pi}{R},
\end{equation}

\noindent where we write explicitly the normalization factor and use $L_z = v T$, where $v$ is a velocity of the incident electron. The large factor $R$ does not cancel here % for the number of events,
because in our setup the wave beam is not regularized (compare with \cite{karlovets2017}). However, it can be approximately identified with the beam size, see Appendix \ref{AppA}. Moreover, as we shall see further, in the physically meaningful results this factor will eventually cancel out.%, so this ambiguity does not raise a problem in our current study.

Substituting the single atom scattering amplitude (\ref{single_amp}) into Eq.~(\ref{noe_single_1}) we find

\begin{equation}
\begin{aligned}
    \left(\frac{\dd \nu}{\dd \Omega}\right)^{(single)} 
    &= \frac{N_e}{\cos{\theta_p}}\frac{\kappa}{R} \, 2Z^2e^4(\varepsilon^2\delta_{\lambda\lambda^\prime} + m_e^2\delta_{\lambda,-\lambda^\prime}) \\
    &\cross \abs{\sum\limits_{\sigma=\pm 1/2} d_{\sigma\lambda}^{1/2}(\theta_p)d_{\sigma\lambda^\prime}^{1/2}(\theta^\prime)\mathcal{I}_{m-\sigma}(\alpha,\beta,\vb{b}_0)}^2 
\end{aligned}
    \label{noe_single_2}
\end{equation}

For small $\theta_p$ this formula is reduced to

\begin{equation}
    \begin{aligned}
    \left(\frac{\dd \nu}{\dd \Omega}\right)^{(single)} 
    &= N_e \frac{\kappa}{2R} J_{m-\lambda}^2(\kappa b_0) \left(\frac{\dd \sigma}{\dd \Omega}\right)^{(PW)} \\
    &= L^{(TW)} J_{m-\lambda}^2(\kappa b_0) \left(\frac{\dd \sigma}{\dd \Omega}\right)^{(PW)}
    \end{aligned}
    \label{noe_single_3}
\end{equation}

Note that for $\kappa = 0$ this quantity equals  to the plane wave cross-section times the luminosity factor. In principle, we can backtrack this relation and define the twisted cross-section for the scattering by a single atom target the following way \cite{karlovets2017}:

\begin{equation}
\begin{aligned}
    \left(\frac{\dd \sigma}{\dd \Omega}\right)^{(single)} &\equiv \frac{1}{L^{(TW)}}\left(\frac{\dd \nu}{\dd \Omega}\right)^{(single)} \\
    &= J_{m-\lambda}^2(\kappa b_0) \left(\frac{\dd \sigma}{\dd \Omega}\right)^{(PW)}
\end{aligned}
\end{equation}

\subsection{Mesoscopic target}

It was shown in \cite{serbo2015} that with a macroscopic target, the sensitivity to the OAM of the incoming twisted electron is lost.
On the other hand, a single atom target is not the simplest experimentally realizable option, though a possible one (see, for example, \cite{Ott_2016}). Moreover, a trapped atom target has a spatial probability distribution of some considerable width (typically hundreds of nanometers) and therefore is realistically described as a wave-packet and the method of the previous subsection is not applicable.
A finite size mesoscopic target appears as a more experimentally viable option, is easier to prepare (with respect to a single atom one) and could still lead to experimental differentiation of the OAM values. We shall model such a target by a Gaussian distribution (\ref{distr_meso}):

%We can try to use a target of a finite size, which would be easier to prepare than a single atom one, and which would still lead to differences in scattering results due to different values of OAM of the twisted electron. 

\begin{equation}
    \abs{F_{m,\lambda,\lambda^\prime}^{(meso)}(\vb{p},\vb{p}^\prime,\vb{b}_0)}^2  =
    \int \dd^2b \,\abs{F(\vb{p},\vb{p}^\prime,\vb{b})}^2 \frac{e^{-\frac{1}{2}\left(\frac{\vb{b}-\vb{b}_0}{\sigma_b}\right)^2}}{2\pi\sigma_b^2}
    \label{meso1}
\end{equation}

The calculations of the amplitude and the number of events are similar to the ones in the previous section and can be found in Appendix \ref{AppC}.

Number of events in the limit of small $\beta$ and $\vb{b}_0 = 0 $ equals 

\begin{equation}
\begin{aligned}
    &\left(\frac{\dd \nu}{\dd \Omega}\right)^{(meso)} = e^{-\sigma_b^2\kappa^2}\frac{2Z^2e^4}{\alpha^2}I_{m-\lambda}(\sigma_b^2\kappa^2)\\
    &\cross \frac{\kappa}{R}  \frac{N_e}{\cos{\theta_p}}\left(\varepsilon^2\cos{(\theta^\prime/2)}\delta_{\lambda,\lambda^\prime}+m_e^2\sin{(\theta^\prime/2)}\delta_{\lambda,-\lambda^\prime}\right) \\
    &= e^{-\sigma_b^2\kappa^2}I_{m-\lambda}(\sigma_b^2\kappa^2) L^{(TW)} \left(\frac{\dd \sigma}{\dd \Omega}\right)^{(PW)}.
\end{aligned}
\label{noe_meso_limit}
\end{equation}

\noindent where $I_{m-\lambda}(\sigma_b^2\kappa^2)$ is the modified Bessel function of 1st kind.

For a point-like target, $\sigma_b = 0$, the modified Bessel function turns out to be equal to
\begin{equation}
    I_{m-\lambda}(0) = \delta_{m-\lambda,0}
\end{equation}
Then, we have the following expression for the number of events:
\begin{equation}
\begin{aligned}
    &\left(\frac{\dd \nu}{\dd \Omega}\right)^{(meso)}
    \rightarrow \delta_{m-\lambda,0} L^{(TW)} \left(\frac{\dd \sigma}{\dd \Omega}\right)^{(PW)} \\
    &= \left(\frac{\dd \nu}{\dd \Omega}\right)^{(single)}(\vb{b}=0)
\end{aligned}
\label{noe_meso_limit_point}
\end{equation}

For a large target with $\sigma_b \sim R \rightarrow \infty$, the modified Bessel function has the following limit:

\begin{equation}
    I_{m-\lambda}(\sigma_b^2\kappa^2) \rightarrow \frac{e^{\sigma_b^2\kappa^2}}{\sqrt{2\pi}\sigma_b\kappa}
    \label{bessel_i_lim}
\end{equation}

Then for the number of events we have:

\begin{equation}
% \begin{aligned}
    \left(\frac{\dd \nu}{\dd \Omega}\right)^{(meso)} = \frac{1}{\sqrt{2\pi}\sigma_b\kappa} \frac{\kappa}{2R}  \frac{N_e}{\cos{\theta_p}}\left(\frac{\dd \sigma}{\dd \Omega}\right)^{(PW)}
% \end{aligned}
\label{noe_meso_limit_large}
\end{equation}

For the realistic atomic potential (\ref{potential_at}) and small $\beta$, we find

\begin{equation}
\begin{aligned}
    &\left(\frac{\dd \nu}{\dd \Omega}\right)^{(meso)} = e^{-\sigma_b^2\kappa^2}2Z^2e^4 \left(\sum\limits_{i=1}^3 \frac{A_i}{\alpha_i}\right)^2 I_{m-\lambda}(\sigma_b^2\kappa^2)\\
    &\cross \frac{\kappa}{R}  \frac{N_e}{\cos{\theta_p}}\left(\varepsilon^2\cos{(\theta^\prime/2)}\delta_{\lambda,\lambda^\prime}+m_e^2\sin{(\theta^\prime/2)}\delta_{\lambda,-\lambda^\prime}\right).
\end{aligned}
\label{noe_meso_limit_large_at}
\end{equation}

To compare the scattering by the mesoscopic target to the macroscopic scenario, we need to introduce $\dd\nu/\dd\Omega$ in the latter case:
%Doing the same steps as for other kinds of targets, we find:
\begin{equation}
    \left(\frac{\dd \nu}{\dd \Omega}\right)^{(macro)} = \frac{N_e}{16\pi^2}\frac{\abs{\vb{p}}}{\varepsilon} \abs{F_{m,\lambda,\lambda^\prime}^{(macro)}(\vb{p},\vb{p}^\prime)}^2.
    \label{noe_macro_1}
\end{equation}
In the limit of small $\theta_p$,
\begin{equation}
    \begin{aligned}
        \left(\frac{\dd \nu}{\dd \Omega}\right)^{(macro)} &= \frac{\pi}{R} \frac{N_e}{\cos{\theta_p}} \frac{1}{\pi^2R} \left(\frac{\dd \sigma}{\dd \Omega}\right)^{(PW)} \\
        &= \frac{2}{\pi R\kappa} \frac{\kappa}{2R} \frac{N_e}{\cos{\theta_p}} \left(\frac{\dd \sigma}{\dd \Omega}\right)^{(PW)}
    \end{aligned}
    \label{noe_macro_2}
\end{equation}

If we assume that $R=2\sqrt{2/\pi}\sigma_b$, then (\ref{noe_macro_2}) is equal to (\ref{noe_meso_limit_large}). This means that in the limit of the large target the scattering of a twisted electron by a mesoscopic target becomes similar to the scattering by a macroscopic one in accordance with our mundane intuition. To “measure” this effect for any given target size $\sigma_b$ we define the following ratio

\begin{equation}
    \mathcal{R}_{m-\lambda}(\sigma_b,\kappa) \equiv \left(\frac{\dd \nu}{\dd \Omega}\right)^{(meso)} {\Big / }\left(\frac{\dd \nu}{\dd \Omega}\right)^{(macro)}
    \label{r_def}
\end{equation}

In the limit of small $\beta$ and for $\vb{b}_0=0$ (typical for relativistic regime) we have

\begin{equation}
    \mathcal{R}_{m-\lambda}(\sigma_b,\kappa) = \sqrt{2\pi}\sigma_b\kappa\, e^{-\sigma_b^2\kappa^2}I_{m-\lambda}(\sigma_b^2\kappa^2)
    \label{r_def_lim}
\end{equation}
Here we used the amplitudes for the potential (\ref{potential}), but this relation holds also for the realistic potential (\ref{potential_at}), since factors $\sum A_i/\alpha_i$ cancel in the considered limit. 

\begin{figure*}[!htb]
    \centering
    \includegraphics[width=\linewidth]{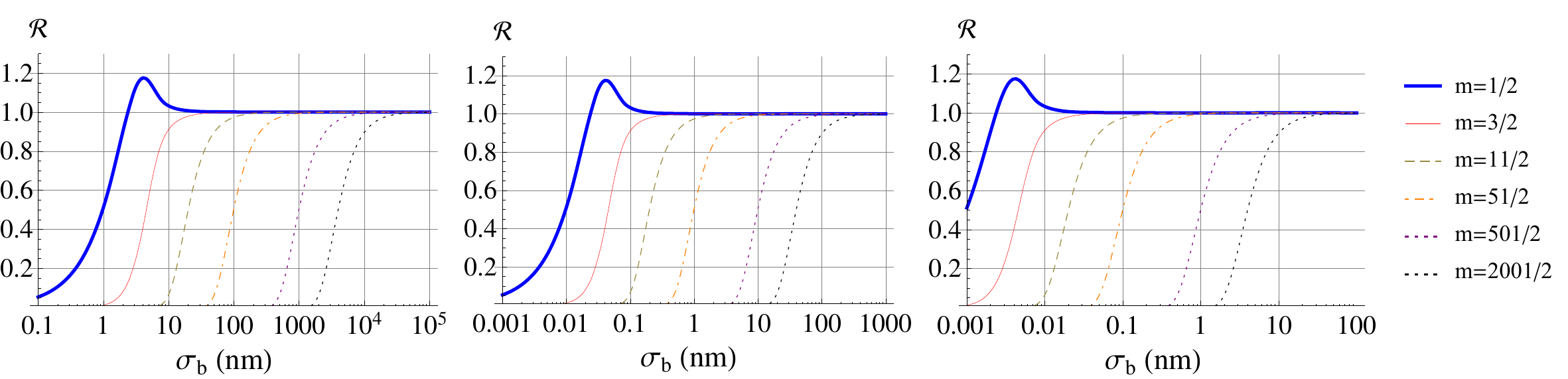}
    \caption{The ratio of the numbers of events $\mathcal{R}_{m-\lambda}= \left(\dd \nu/\dd \Omega\right)^{(meso)} / \left(\dd \nu/\dd \Omega\right)^{(macro)}$ from Eq.~(\ref{r_def_lim}) for different TAM projections $m$. Parameters: $\varepsilon = 5\,m_e$, $\lambda = 1/2$. \textit{Left panel}: $\theta_p = 0.001$ deg, $\kappa$ = 44 eV; \textit{middle panel}: $\theta_p = 0.1$ deg, $\kappa$ = 4.4 keV; \textit{right panel}: $\theta_p = 1$ deg, $\kappa$ = 44 keV.
    }
    \label{fig:meso1}
\end{figure*}

Our idea behind the introduction of the function $\mathcal{R}$ is that it would manifest how the scattering result varies with changing the target size from a point-like atom to an infinitely large target. Several examples are given in Figure \ref{fig:meso1}. 
Moreover, with this ratio we can explore the process sensitivity to the incident electron OAM value.
%It is evident from the figure that for different values of m, the transition region position varies. %Moreover, with this ratio we can explore the process sensitivity to the incident electron OAM value change with varying the target size.

So far we used the Gaussian distribution (\ref{distr_meso}) to model the mesoscopic target. Alternatively, we can use the uniform distribution on a finite interval: 
\begin{equation}
    \abs{F_{m,\lambda,\lambda^\prime}^{(meso)}(\vb{p},\vb{p}^\prime,0)}^2 =
    \int\limits_{S_b} \dd^2b \,\abs{F(\vb{p},\vb{p}^\prime,\vb{b})}^2 \frac{1}{\pi R_b^2}  
\end{equation}

\noindent where $R_b$ is a radius of the circular target and $S_b$ is its area. In terms of the number of events, we have the following:

\begin{equation}
    \left(\frac{\dd \nu}{\dd \Omega}\right)^{(meso)} =
    \int\limits_{S_b} \dd^2b \,\frac{1}{\pi R_b^2}\left(\frac{\dd \nu}{\dd \Omega}\right)^{(single)}  , 
    \label{meso_uniform}
\end{equation}

Eq.~(\ref{noe_single_3}) can be used for the single atom target in the limit $\beta \rightarrow 0$, then for Eq.~(\ref{meso_uniform}) we find:

\begin{equation}
    \left(\frac{\dd \nu}{\dd \Omega}\right)^{(meso)} =
    \int\limits_{S_b} \dd^2b \,\frac{1}{\pi R_b^2}J_{m-\lambda}^2(\kappa b) L^{(TW)}\left(\frac{\dd \sigma}{\dd \Omega}\right)^{(PW)}.
\end{equation}

\begin{figure}[!htb]
    \centering
    \includegraphics[width=0.65\linewidth]{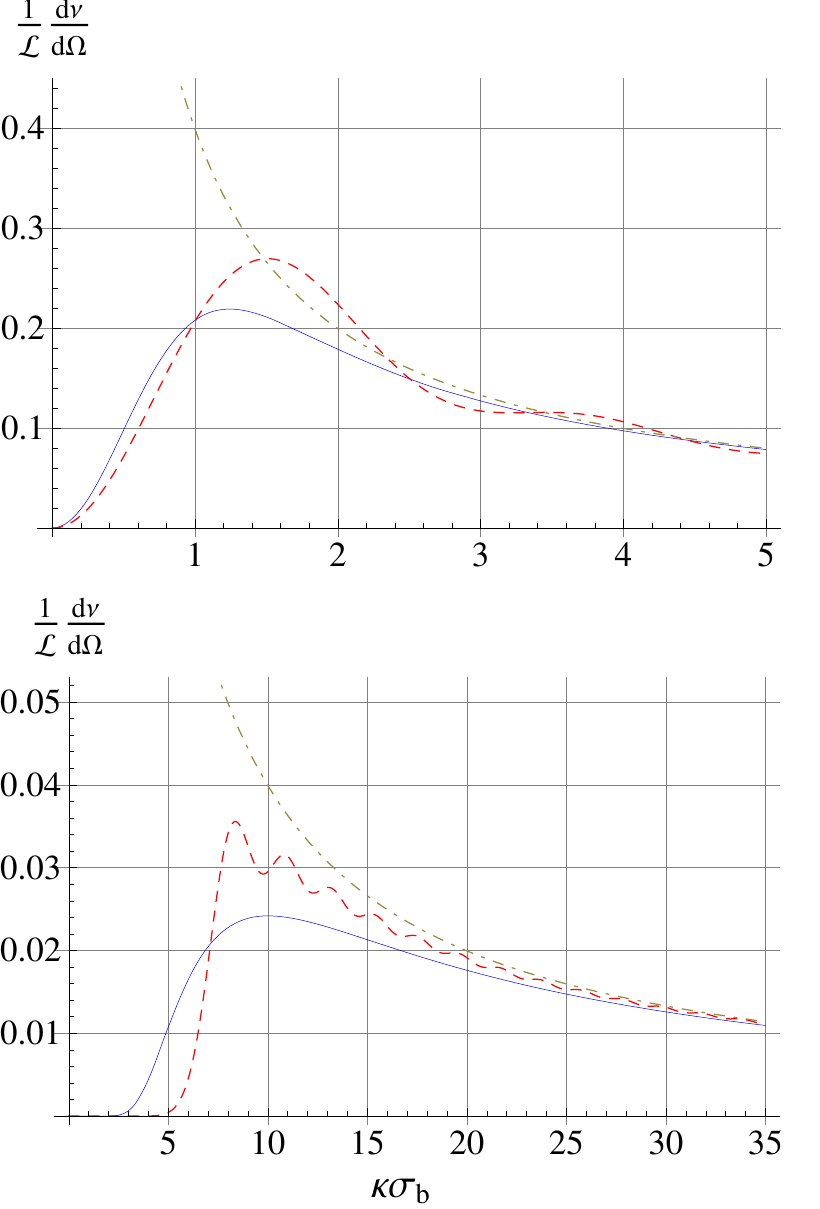}
    \caption{Comparison of the number of scattering events functions for the Gaussian mesoscopic (blue solid line), the uniform mesoscopic (red dashed line), and the macroscopic (green dot dashed line) targets. Parameters: $m-\lambda = 1$ (\textit{top panel}) and $m-\lambda = 10$ (\textit{bottom panel}), the uniform target radius $R_b = 2\sqrt{2/\pi}\sigma_b$, see before Eq.~(\ref{r_def}).
    }
    \label{fig:uniform}
\end{figure}

This integral can be evaluated numerically, and the results are shown in Figure \ref{fig:uniform}, where we plot $(\dd\nu/\dd\Omega)/\mathcal{L}$,  with $\mathcal{L}\equiv L^{(TW)}\cdot(\dd\sigma/\dd\Omega)^{(PW)}$, for the Gaussian mesoscopic (\ref{noe_meso_limit}), the uniform mesoscopic and  the macroscopic targets, all in the relativistic limit of small $\beta$.

\section{Results}\label{section_iv}

%In the previous sections, we have derived the cross-section for the scattering of the twisted electron by a macroscopic (infinitely wide) target and the number of events for scattering by a mesoscopic (target with a finite size $\sigma_b$) and by a single atom. Since we are interested mainly in the realistic case when for relativistic regime transverse momentum $\kappa$ is small relative to total linear momentum, we used this approximation to obtain analytical results, which is complicated in the general case, as mentioned above. For the mesoscopic target, we also assumed that the target is centered ($\vb{b} = 0$).

First, we would like to numerically motivate an approximation used in the previous section, i.e. that for the relativistic regime one can assume the transverse momentum $\kappa$ to be small relative to total linear momentum of the incident electron.
We start with determining the opening angle $\theta_p$ value from the typical scale of the incident wave beam.
In our framework we can define the characteristic beam width $r_{beam}$ either from the half-width radius of the squared wave function for the zero-order beam or 
from matching the first maximum for the higher-order modes (first maximum of $J_n(z)$ is situated at $z \approx n$): 
%Therefore, we have for the transverse momentum the following expression:

\begin{equation}
    \kappa \approx 
    \left\lbrace \begin{aligned}
    &1/r_{beam}, \text{ if $m - \lambda = 0$} \\
    &\frac{m - \lambda}{r_{beam}}, \text{ 
\,\,\,\,otherwise}
    \end{aligned}\right.
    \label{kappa_approx}
\end{equation}

\noindent For an angle between the total linear momentum and the propagation axis that gives:

\begin{equation}
    \theta_p \approx 
    \left\lbrace \begin{aligned}
    &\arcsin{\left(\frac{1}{r_{beam}\sqrt{\varepsilon^2 - m_e^2}}\right)}, \text{ if $m - \lambda = 0$} \\
    &\arcsin{\left(\frac{m - \lambda}{r_{beam}\sqrt{\varepsilon^2 - m_e^2}}\right)}, \text{ 
otherwise.}
    \end{aligned}\right.  
\end{equation}

In Table \ref{tab:table1} we assemble our estimations for $\kappa$ and $\theta_p$ for various widths $r_{beam}$ after setting $\varepsilon = 2m_e$. Such beam widths are currently obtainable for the electron vortices  \cite{Verbeeck2011, SCHATTSCHNEIDER201481}.
We can see that small values of $\theta_p$ ($\lesssim1$~deg) are typical for a rather wide range of beam parameters, thus justifying the approximation made in the previous section.
In case of larger incident energy $\varepsilon$ and other parameters fixed, $\theta_p$ tends to become even smaller, as $\theta_p \propto \varepsilon^{-1}$ for large $\varepsilon$.
%We see there that typical values of the angle $\theta_p$ are smaller or of the order of 1 degree, so the small angle approximation is valid for our problem. For higher energies of the electron for a fixed size of the beam, the angle $\theta_p$ will be even smaller. For large $\varepsilon$ we have $\theta_p \propto \varepsilon^{-1}$. 
For example, for $\varepsilon = 10m_e$, $m-\lambda = 1$ and $r_{beam} = 1$ nm we have $\theta_p=1.15\cross 10^{-3}$ deg. 
We remark, though, that large $\theta_p$ values are not uncommon for some energies, see for instance the analysis of \cite{serbo2015} for the non-relativistic incident electron.
%Larger angle values could be obtained for smaller energies of the electron, but this is the case of non-relativistic energies, which was studied in \cite{serbo2015}.

\begin{table*}[!htb]
\caption{\label{tab:table1}%
The values of the transverse momentum 
 $\kappa$ (eV) and the angle $\theta_p$ (deg) with $\varepsilon = 2m_e$ and different wave beam width.
}
\begin{ruledtabular}
\begin{tabular}{c|c|c|c|c}
% \colrule
 &\multicolumn{4}{c}{$\kappa$ (eV) ($\theta_p$ (deg))}\\
\cline{2-5}
\textrm{$m - \lambda$} & $r_{\textbf{beam}}$ = 1 \r{A} & $r_{\textbf{beam}}$ = 1 nm & $r_{\textbf{beam}}$ = 10 nm & $r_{\textbf{beam}}$ = 1 \textmu m \\
\hline
1 & 2$\cross 10^3$ (1.3$\cross 10^{-1}$) & 2$\cross 10^2$ (1.3$\cross 10^{-2}$) & 2$\cross 10$ (1.3$\cross 10^{-3}$) & 2$\cross 10^{-1}$ (1.3$\cross 10^{-5}$) \\
5 & 1$\cross 10^4$ (6.6$\cross 10^{-1}$) & 1$\cross 10^3$ (6.6$\cross 10^{-2}$) & 1$\cross 10^2$ (6.6$\cross 10^{-3}$) & 1 (6.6$\cross 10^{-5}$) \\
10 & 2$\cross 10^4$ (1.3) & 2$\cross 10^3$ (1.3$\cross 10^{-1}$) & 2$\cross 10^2$ (1.3$\cross 10^{-2}$) & 2 (1.3$\cross 10^{-4}$) \\
100 & 2$\cross 10^5$ (1.3$\cross 10$) & 2$\cross 10^4$ (1.3) & 2$\cross 10^3$ (1.3$\cross 10^{-1}$) & 2$\cross 10$ (1.3$\cross 10^{-3}$) \\
1000 & -- & 2$\cross 10^5$ (1.3$\cross 10$) & 2$\cross 10^4$ (1.3) & 2$\cross 10^2$ (1.3$\cross 10^{-3}$) \\
\end{tabular}
\end{ruledtabular}
\end{table*}

\begin{figure}[!htb]
    \centering
    \includegraphics[width=1\linewidth]{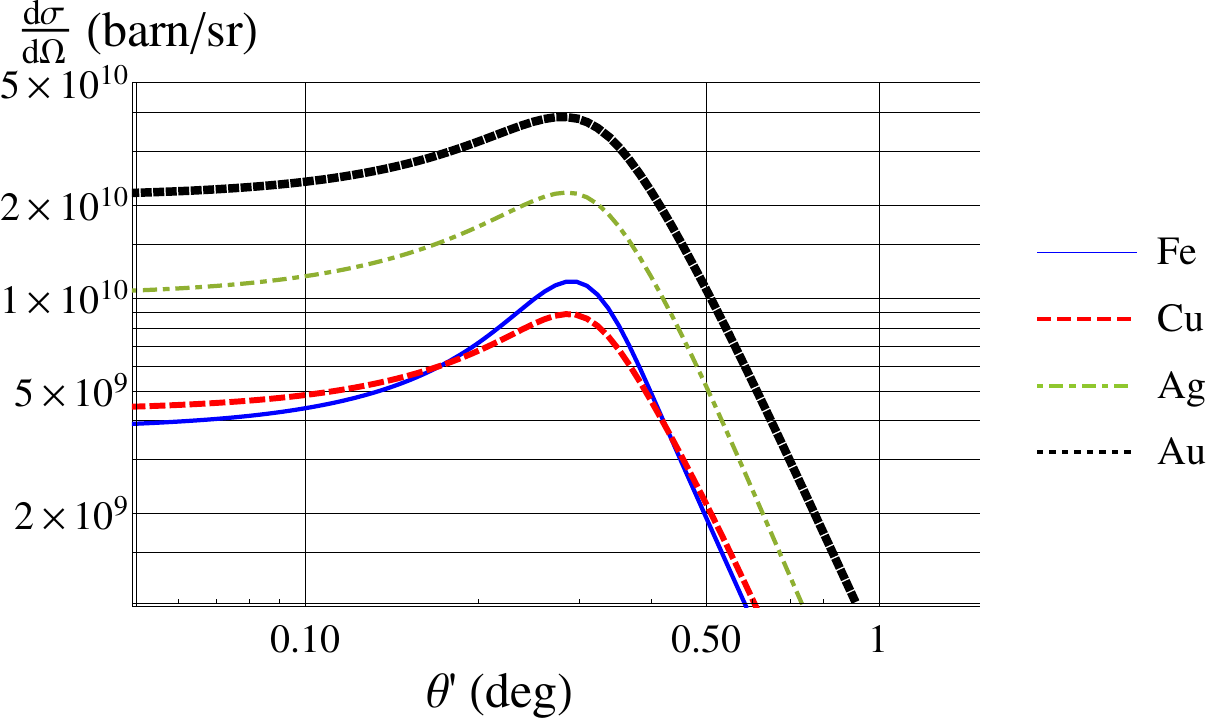}
    \caption{The cross-section for the macroscopic target made of different elements -- iron, copper, silver and gold. Parameters: $\varepsilon = 5m_e$, $\theta_p = 0.3^\circ$ ($\kappa = 13.1$ keV).
    }
    \label{fig:macro_dif}
\end{figure}

Let us now go back to the figures that were introduced earlier in the paper.
In Figure \ref{fig:macro_fe} the results for differential cross-section for incident relativistic ($\varepsilon/m_e=2,\ 5,\ 20$) twisted electron scattered by a macroscopic iron target are presented. The position of the peak corresponds to the value of the opening angle for the twisted electron $\theta^\prime = \theta_p$; determining $\theta^\prime$ from the scattering and knowing the electron energy from beforehand -- one can easily calculate the transverse momentum $\kappa$. However, this peak is distinguishable only for values of $\kappa \gtrsim 10$ keV. This sets a threshold for measuring the $\kappa$ with the described method. The value of 10 keV comes from the screening parameter $\mu$, which is approximately equal to the inverse Bohr radius, being the natural scale of the problem. In Figure \ref{fig:macro_au_he} we take a golden macroscopic target and electrons with ultra-relativistic energies ($\varepsilon/m_e=100,\ 1000$) and observe a similar picture. In fact, the scattering picture does not alter substantially with the change of the element -- for higher values of $Z$ the cross-section increases in general, but the peak becomes less pronounced, as can be seen in Figure \ref{fig:macro_dif} for iron, copper, silver, and gold targets. 

The values of $\kappa \gg 10$~keV in Figures \ref{fig:macro_fe}, \ref{fig:macro_au_he} are hardly achievable in experiment and are presented here mainly to illustrate the tendencies of the cross-sections. Furthermore, for $\kappa \gg 10$~keV the width of the beam may became smaller than size of an atom, and that can make our target model inapplicable. However, increase of $\kappa$ can be compensated by the increase of the OAM value, see Eq. (\ref{kappa_approx}).

The scattering by a macroscopic target is at a disadvantage because it can not provide information about the electron OAM, whereas the scattering by a single atom  target or by a mesoscopic target can. We reproduce here the formula for the single atom scattering amplitude (\ref{noe_single_2}) and find its approximation for small angle $\theta_p$ (\ref{noe_single_3}). In  the relativistic regime, the amplitude of this process can be found analytically. %The single atom scattering amplitude is given in (\ref{noe_single_2}) for the general case and in (\ref{noe_single_3}) for a small $\theta_p$.

For the scattering by a mesoscopic target, we obtain an analytical solution for the amplitude when the target is centered ($\vb{b}_0=0$) and the angle $\theta_p$ is small.  This approach was studied in \cite{karlovets2017} for the non-relativistic regime, where the Gaussian distribution was also used for modeling the mesoscopic target. We show that for the limit of small or large target size $\sigma_b$, this amplitude reduces to the single atom (\ref{single_amp}) or the macroscopic (\ref{noe_macro_2}) scattering amplitudes, respectively. By comparing the mesoscopic and the macroscopic amplitudes, we explore the transition between the mesoscopic and the macroscopic target scattering behavior and observe how the sensitivity to the angular momentum disappears. This phenomenon can be used to estimate the twisted electron orbital angular momentum in experiment.

In Figure \ref{fig:meso1} we plot the ratio (\ref{r_def_lim}) as a function of the target size $\sigma_b$ for different transverse momentum $\kappa$ of the incident electron. The curve corresponding to $m-\lambda = 0$ is different from the others because the zero-order Bessel function behaves distinctly. Since $J_{m-\lambda}(0) = \delta_{m-\lambda,0}$, the wave function with the zero OAM interacts with a small target in its center, while for non-zero values of the OAM, the target “goes through” the wave beam center without overlapping  its probability density. In (\ref{r_def_lim}) this can also be seen directly because the modified Bessel function with small argument behaves similarly when in it has zero index.

In Figure \ref{fig:meso1}, for every curve we can observe an evolution from left (small target size) to  right (wide target) as an electron beam goes from the “single atom” scattering scenario, through the mesoscopic scattering and, finally, to the macroscopic scattering scenario. It can be seen from the single atom scattering (\ref{noe_single_3}) that, for $m - \lambda \neq 0$ and when the target is on the propagation axis (impact parameter $\vb{b}_0 = 0$), the scattering amplitude is equal to zero, that explains the smaller values on the left-side of the curves. For the macroscopic target, there is no dependence on the orbital angular momentum $m - \lambda$, so all curves converge to unity in the right part of this figure.

\section{Discussion and conclusion}\label{section_v}

% We explore here the use of scattering by a screened Coulomb potential to analyze properties of the twisted relativistic electron beams, such as its transverse momentum $\kappa$ (and so its momentum cone angle $\theta_p$) and its orbital angular momentum $m-\lambda$.
% %{\color{blue} In this paper we studied the scattering of the twisted relativistic electron by a screened Coulomb potential to analyze such properties of the electron as its transverse momentum $\kappa$ (closely related to its momentum cone angle $\theta_p$) and its orbital angular momentum $m-\lambda$.} 
% We have shown that using a macroscopic target allows one to measure $\theta_p$ if the transverse momentum has values of at least $10$ keV, otherwise the corresponding resonance peak would not be distinguishable. The electrons with such a high transverse momentum can in principle be generated via scattering processes at accelerator facilities, especially when employing the generalized measurement technique \cite{Karlovets2022_gen}.

In this paper, we have shown how the scattering by a screened Coulomb potential can be used to analyze the properties of the twisted relativistic electron beams. The scattering by a macroscopic target allows one to measure the transverse momentum $\kappa$ (and the beam cone angle $\theta_p$), under a condition that $\kappa$ has values of at least $10$ keV. The electrons with such high transverse momenta can in principle be generated via scattering processes at accelerator facilities, especially when employing the generalized measurement technique \cite{Karlovets2022_gen}.

Moreover, we have demonstrated how a target of a finite size (\textit{mesoscopic}) can be used to retrieve information about the twisted electron orbital angular momentum $m-\lambda$. 
Our method allows one to estimate the electron OAM by taking targets of different sizes $\sigma_b$ and analyzing the ratio $\mathcal{R}_{m-\lambda}(\sigma_b,\kappa)$ of the number of events for mesoscopic and macroscopic targets, assuming the electron transverse momentum $\kappa$ is known. 
% By taking targets of different sizes and by observing and comparing the scattering results (the number of events) for a macroscopic target and for a mesoscopic one, it is possible to estimate the electron OAM when the transverse momentum $\kappa$ is known.
For $\kappa \sim 40$ eV, the OAM of any value can be retrieved for realistic targets wider than $1$ nm. In contrast, for higher values of $\kappa$, the scattering process is sensitive only to higher values of the OAM for such targets -- {\it e.g.}, for $\kappa = 4.4$ keV, one can distinguish $m-\lambda$ starting from $50$ and higher. 
%In this particular case, the proposed method does not discern OAMs outside the range $\sim 50\pm 10$
For $\kappa = 44$ keV, the lowest retrievable value of OAM is $m-\lambda = 500$. In general, for high enough transverse momentum ($\kappa \gtrsim 1$~keV) there is a restrictive bottom bound in the range of measurable OAM values for the realistic target sizes ($\gtrsim 1$ nm). Increasing $\kappa$ leads to a rise of this bottom bound value. On the other hand, the proposed method  does not in principle impose an upper bound on the OAM value we can possibly measure.

To conclude, we estimate the sensitivity of the proposed OAM measurement method.
Let us consider two values of the OAM $m_2-\lambda_2$ and $m_1-\lambda_1$, which we wish to distinguish, and introduce the following relations

\begin{align}
    &\delta = \frac{(m_2-\lambda_2)-(m_1-\lambda_1)}{m_2-\lambda_2}, \\
    &\mathcal{D} = \max_{\sigma_b,\kappa \in \mathbb{R}_+} \left(\mathcal{R}_{m_2-\lambda_2}(\sigma_b,\kappa)-\mathcal{R}_{m_1-\lambda_1}(\sigma_b,\kappa)\right),
\end{align}

\noindent where $m_2-\lambda_2 > m_1-\lambda_1$. The relation $\delta$ characterizes the OAM detuning and the function $\mathcal{D}$ quantifies the necessary accuracy in the scattering amplitude measurement.
If there is a two times difference between the OAM values ($\delta = 0.5$) we have $\mathcal{D} \approx 0.45$. For closer OAM values (smaller $\delta$), $\mathcal{D}$ also decreases: for $\delta \approx 0.08$ we have $\mathcal{D} \approx 0.064$, and for $\delta \approx 0.01$ we have $\mathcal{D} \approx 0.007$.
For example, to distinguish $m_1-\lambda_1=11$ and $m_2-\lambda_2=12$ (detuning $\delta = 0.083$) one must have experimental setup resolution better than  $\mathcal{D} = 0.064$.
We notice that  $\mathcal{D}$ and $\delta$ are of the same order of magnitude. One can use this fact at a preliminary stage of experiment planning.

\begin{acknowledgments}
We are grateful to S. Baturin, G. Sizykh, D. Grosman, N. Sheremet and I. Pavlov for fruitful discussions and criticism. The studies in Sec.\,II are supported by the Government of the Russian Federation through the ITMO Fellowship and Professorship Program and by the Foundation for the Advancement of Theoretical Physics and Mathematics “BASIS”. %The studies in Sec.\,III are supported by the Russian Science Foundation (Project No.\,21-42-04412; https://rscf.ru/en/project/21-42-04412/). 
The studies in Sec.\,III are supported by the Ministry
of Science and Higher Education of the Russian Federation (agreement No. 075-15-2021-1349).
The studies in Sec.\,IV are supported by the Russian Science Foundation (Project No.\,23-62-10026;  https://rscf.ru/en/project/23-62-10026/).
\end{acknowledgments}

\begin{appendix}

\section{Normalization of wave functions}\label{AppA}

%First, let us look at the plane wave normalization. We can write it down next way
The plane wave normalization is introduced as follows
\begin{equation}
    \psi^{PW} = N^{PW}u_{\vb{p}\lambda}e^{ipx}.
\end{equation}
The wave function should obey a normalization condition on $\rho = j_0$, $\int_V \dd^3 r \, \rho(\vb{r})=1$.
That results in
\begin{equation}
    N^{PW} = \frac{1}{\sqrt{2\varepsilon V}}.
\end{equation}

Let us follow the same steps to find the normalization constant of the twisted wave function,
\begin{equation}
    \psi_{\kappa m p_z \lambda}(\vb{r}) = N^{TW}\int \frac{\dd^2 p_\perp}{(2\pi)^2}a_{\kappa m}(\vb{p}_\perp)\psi_{\vb{p} \lambda},
\end{equation}
Using  $\int\limits_0^R J_n^2(\kappa r)\kappa r \dd r = \frac{R}{\pi}$ (\cite{Jentschura_2011}), we find

\begin{equation}
    N^{TW} = \sqrt{\frac{\pi}{2\varepsilon RL_z}}.
    \label{twisted_norm}
\end{equation}

We could also regularize the twisted wave function density employing the Gaussian distribution $\exp{-1/2(r/\sigma)^2}$. Then we have 

\begin{equation}
    \kappa\int\limits_0^R J_n^2(\kappa r) e^{-\frac{1}{2}\left(\frac{r}{\sigma}\right)^2}\,r \dd r = 2\pi\sigma^2\kappa e^{-\sigma^2\kappa^2}I_n(\sigma^2\kappa^2)
\end{equation}

For a large value of $\sigma$ there is a limit (see Eq.~(\ref{bessel_i_lim})) for the Bessel function reducing the expression above  to $\sqrt{2\pi}\sigma$.
In this case the normalization constant is

\begin{equation}
    N^{TW,reg} = \sqrt{\frac{1}{2\varepsilon \sqrt{2\pi}\sigma L_z}}
\end{equation}

Therefore, we can interpret $R$ in (\ref{twisted_norm}) as a Bessel beam width.

\section{Evaluation of the integrals with method of stationary phase}\label{AppB}

For small $\kappa b$, the integral (\ref{int_def}) can be easily evaluated numerically. However, for large $\kappa b$ the integral becomes highly oscillatory making the calculation much harder. On the other hand, the method of stationary phase \cite{fedoryuk, bender1999} is applicable in this limit. This method is used to evaluate integrals of the following form:

\begin{equation}
    F(\lambda) = \int\limits_a^b f(x) \exp[i\lambda S(x)]\dd x,
    \label{stat_def1}
\end{equation}

\noindent where $\lambda \gg 1$ is a large parameter. The point $x_0$ where $S' (x) = 0$ is called a stationary point. If ($S''(x_0)\neq 0$), we have the following approximation

% \begin{equation}
%     \begin{aligned}
%     F(\lambda;x_0) &\equiv \int\limits_{x_0-\delta}^{x_0+\delta} f(x) \exp[i\lambda S(x)]\dd x \sim \\&\exp\left[i\lambda S(x_0) +\frac{i\pi}{4}\text{sign }S''(x_0)\right]\lambda^{-\frac{1}{2}} \sum\limits_{k=0}^\infty a_k \lambda^{-k}   
%     \end{aligned}
% \end{equation}

% for main asymptotics term

\begin{equation}
    \begin{aligned}
    F(\lambda;x_0) = \sqrt{\frac{2\pi}{\lambda\abs{S''(x_0)}}}\left[f(x_0) + O(\lambda^{-1})\right] \\
    \cross \exp\left[i\lambda S(x_0) +\frac{i\pi}{4}\text{sign }S''(x_0)\right]
    \end{aligned}
    \label{stat_def2}
\end{equation}

Let us see how this method can be applied to the integral (\ref{int_def}) (in principle, it also can be used for (\ref{int_def2})). In terms of (\ref{stat_def1}) the parameters are $\lambda = \kappa b$, $x = \phi$. If $n\ll \kappa b$, then we can take

\begin{align}
    S(x) &= - \cos{(\phi+\varphi^\prime-\varphi_b)},\\ f(x) &= \frac{1}{2\pi}\frac{e^{in\phi}}{\alpha-\beta\cos{\phi}}
\end{align}

Here $S'(x) = \sin{(\phi+\varphi^\prime-\varphi_b)}$ and it equals to zero if $\phi = \varphi_b-\varphi^\prime$ or $\phi = \varphi_b-\varphi^\prime + \pi$, thus we have two stationary points. Using (\ref{stat_def2}), we find the stationary phase approximation for Eq.~(\ref{int_def}):

\begin{equation}
    \begin{aligned}
    \mathcal{I}_{n}(\alpha,\beta,\vb{b}) &\approx \\ \sqrt{\frac{1}{2\pi\kappa b}}&\left[\frac{e^{in(\varphi_b-\varphi^\prime)}e^{-i(\kappa b -\frac{\pi}{4})}}{\alpha-\beta\cos{(\varphi_b-\varphi^\prime)}} -\frac{e^{in(\varphi_b-\varphi^\prime)}e^{i(\kappa b -\frac{\pi}{4})}}{\alpha+\beta\cos{(\varphi_b-\varphi^\prime)}} \right]    
    \end{aligned}
\end{equation}

\begin{figure}[!htb]
    \centering
    \includegraphics[width=0.8\linewidth]{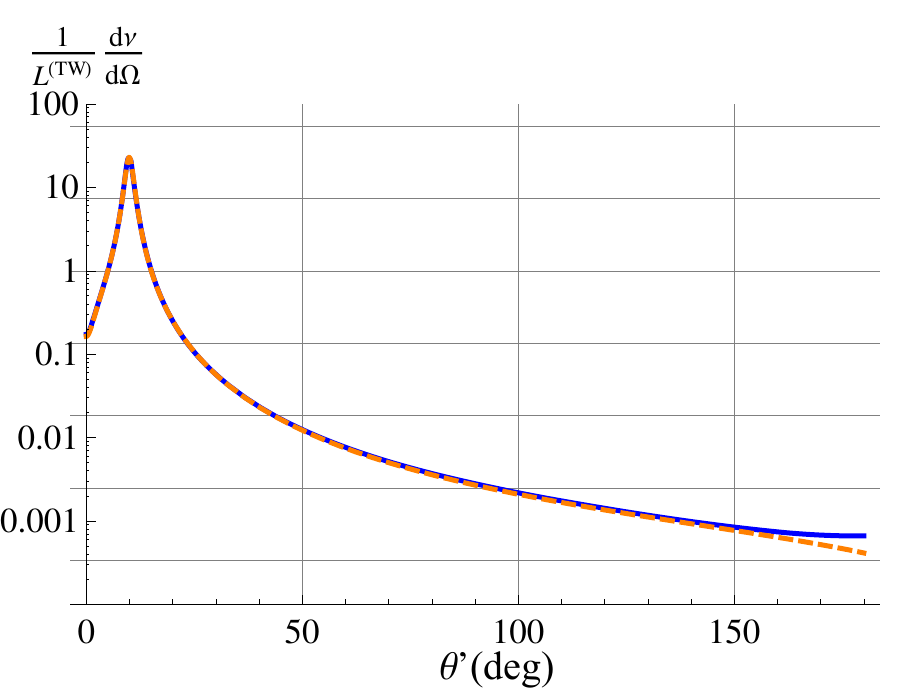}
    \caption{The number of events for the hydrogen single atom target. $\varepsilon = 2m_e$, $\theta_p = 10^\circ$ ($\kappa = 154$ keV), $b = 1$ nm, $m=3/2$, $\lambda = 1/2$. \textit{Blue solid line:} numerical calculation, \textit{orange dashed line:} method of stationary phase.
    }
    \label{fig:stat_comparison}
\end{figure}

In Figure \ref{fig:stat_comparison} we compare a straightforward numerical calculation and the stationary phase approximation for the calculation of the number of events for a single atom target (\ref{noe_single_2}). We see, that the curves  coincide everywhere except the regions where the function becomes too small. The accuracy of the method can be roughly estimated as $(\kappa b)^{-1}$. To obtain higher accuracy one can take the next terms in approximation series, see, for example, \cite{fedoryuk}.

If $n\sim \kappa b$, then $e^{in\phi}$ also oscillates fast, so we should take the following relations instead:

\begin{align}
    S(x) &= \frac{n}{\kappa b} - \cos{(\phi+\varphi^\prime-\varphi_b)},\\ f(x) &= \frac{1}{2\pi}\frac{1}{\alpha-\beta\cos{\phi}}
\end{align}

The next steps are obvious. Note, however, the points where $S''(x) = 0$ in which another formula for the method of stationary phase should be used (see \cite{fedoryuk, bender1999}).

\section{Formulae on the mesoscopic target scattering}\label{AppC}

We present below calculations of mesoscopic scattering amplitude and number of events. The full representation of Eq.~(\ref{meso1}) is

\begin{equation}
    \begin{aligned}
    \abs{F_{m,\lambda,\lambda^\prime}^{(meso)}(\vb{p},\vb{p}^\prime,\vb{b}_0)}^2 &=
    \int \dd^2b \,\abs{F(\vb{p},\vb{p}^\prime,\vb{b})}^2 \frac{e^{-\frac{1}{2}\left(\frac{\vb{b}-\vb{b}_0}{\sigma_b}\right)^2}}{2\pi\sigma_b^2}  \\
    =  \int& \frac{\dd^2 k_\perp}{(2\pi)^2} \frac{\dd^2 p_\perp}{(2\pi)^2} 
 \dd^2 b \, f_{\lambda,\lambda^\prime}(\vb{p},\vb{p}^\prime)f_{\lambda,\lambda^\prime}^*(\vb{k},\vb{p}^\prime) \\ 
    \cross a_{\kappa m}&(\vb{p}_\perp)a_{\kappa m}^*(\vb{k}_\perp)e^{i(\vb{k}_\perp-\vb{p}_\perp)\vb{b}}\frac{e^{-\frac{1}{2}\left(\frac{\vb{b}-\vb{b}_0}{\sigma}\right)^2}}{2\pi\sigma_b^2}.
    \end{aligned}
\end{equation}
The twisted amplitude (\ref{twisted_amp}) can be iserted here straightforwardly, but it is more instructive to expand the equation above as follows:

% \begin{widetext}
% \begin{equation}
%     \begin{aligned}
%     \int&\abs{F^{(m)}_{\lambda,\lambda^\prime}(\vb{p},\vb{p}^\prime,\vb{b}_0)}^2 \frac{1}{2\pi\sigma_b^2} e^{-\frac{1}{2}\left(\frac{\vb{b}-\vb{b}_0}{\sigma_b}\right)^2}\dd^2 b =\\
%     &=\frac{\kappa}{2\pi}\int \frac{\dd\varphi_k}{2\pi} \frac{\dd\varphi_p}{2\pi}e^{im(\varphi_p-\varphi_k)}f_{\lambda,\lambda^\prime}(\vb{p}_\kappa,\vb{p}^\prime)f_{\lambda,\lambda^\prime}^*(\vb{k}_\kappa,\vb{p}^\prime)
%     e^{-\sigma^2\kappa^2(1 - \cos{(\varphi_k-\varphi_p)})}e^{i\kappa\abs{\vb{b}_0}\cos{(\varphi_k-\varphi_b)}}e^{-i\kappa\abs{\vb{b}_0}\cos{(\varphi_p-\varphi_b)}} \\
%     &=\frac{2Z^2e^4}{\pi} \kappa \left(\varepsilon^2\delta_{\lambda,\lambda^\prime}+m_e^2\delta_{\lambda,-\lambda^\prime}\right) e^{-\sigma_b^2\kappa^2} 
%     \sum\limits_{\sigma,\sigma^\prime = -1/2}^{1/2} d_{\sigma,\lambda}^{1/2}(\theta_p)d_{\sigma,\lambda^\prime}^{1/2}(\theta^\prime)d_{\sigma^\prime,\lambda}^{1/2}(\theta_p)d_{\sigma^\prime,\lambda^\prime}^{1/2}(\theta^\prime)\mathcal{I}_{m\sigma\sigma^\prime}^{\sigma_b}(\alpha,\beta,\vb{b}_0),
%     \end{aligned}
%     \label{amp_meso}
% \end{equation}
% \end{widetext}

\begin{equation}
    \begin{aligned}
    \int&\abs{F^{(m)}_{\lambda,\lambda^\prime}(\vb{p},\vb{p}^\prime,\vb{b})}^2 \frac{1}{2\pi\sigma_b^2} e^{-\frac{1}{2}\left(\frac{\vb{b}-\vb{b}_0}{\sigma_b}\right)^2}\dd^2 b \\
    &=\frac{\kappa}{2\pi}\int \frac{\dd\varphi_k}{2\pi} \frac{\dd\varphi_p}{2\pi}e^{im(\varphi_p-\varphi_k)}f_{\lambda,\lambda^\prime}(\vb{p}_\kappa,\vb{p}^\prime) f_{\lambda,\lambda^\prime}^* (\vb{k}_\kappa,\vb{p}^\prime) \\
    &\cross 
    e^{-\sigma^2\kappa^2(1 - \cos{(\varphi_k-\varphi_p)})}e^{i\kappa\abs{\vb{b}_0}\cos{(\varphi_k-\varphi_b)}} \\
    &\cross e^{-i\kappa\abs{\vb{b}_0}\cos{(\varphi_p-\varphi_b)}} =\frac{2Z^2e^4}{\pi} \kappa \left(\varepsilon^2\delta_{\lambda,\lambda^\prime}+m_e^2\delta_{\lambda,-\lambda^\prime}\right)  \\
    &\cross e^{-\sigma_b^2\kappa^2}\sum\limits_{\sigma,\sigma^\prime = \pm 1/2} d_{\sigma,\lambda}^{1/2}(\theta_p)d_{\sigma,\lambda^\prime}^{1/2}(\theta^\prime)d_{\sigma^\prime,\lambda}^{1/2}(\theta_p)d_{\sigma^\prime,\lambda^\prime}^{1/2}(\theta^\prime) \\
    &\cross \mathcal{I}_{m\sigma\sigma^\prime}^{\sigma_b}(\alpha,\beta,\vb{b}_0),
    \end{aligned}
    \label{amp_meso}
\end{equation}

\noindent where we used the property that the Fourier transform of a Gaussian is a Gaussian itself and introduced

% \begin{widetext}
\begin{equation}
\begin{aligned}
    &\mathcal{I}_{m\sigma\sigma^\prime}^{\sigma_b}(\alpha,\beta,\vb{b}_0)  \\
    & =\int \frac{\dd\varphi_p}{2\pi}\frac{e^{im\varphi_p}e^{i\sigma(\varphi^\prime-\varphi_p)}e^{-i\kappa\abs{\vb{b}_0}\cos{(\varphi_p-\varphi_b)}}}{\alpha - \beta\cos{(\varphi_p-\varphi^\prime)}} \\ &\left(\int \frac{\dd\varphi_k}{2\pi}\frac{e^{im\varphi_k}e^{i\sigma^\prime(\varphi^\prime-\varphi_k)}e^{-i\kappa\abs{\vb{b}_0}\cos{(\varphi_k-\varphi_b)}}}{\alpha - \beta\cos{(\varphi_k-\varphi^\prime)}}\right. \\
    &\cross \left. e^{\sigma_b^2\kappa^2 \cos{(\varphi_k-\varphi_p)}} \right)^*.
    \label{int_def2}
\end{aligned}  
\end{equation}
% \end{widetext}

% \begin{widetext}
% \begin{equation}
% \begin{aligned}
%     &\mathcal{I}_{m\sigma\sigma^\prime}^{\sigma_b}(\alpha,\beta,\vb{b}_0) = \\
%     & =\int \frac{\dd\varphi_p}{2\pi}\frac{e^{im\varphi_p}e^{i\sigma(\varphi^\prime-\varphi_p)}e^{-i\kappa\abs{\vb{b}_0}\cos{(\varphi_p-\varphi_b)}}}{\alpha - \beta\cos{(\varphi_p-\varphi^\prime)}} \left(\int \frac{\dd\varphi_k}{2\pi}\frac{e^{im\varphi_k}e^{i\sigma^\prime(\varphi^\prime-\varphi_k)}e^{-i\kappa\abs{\vb{b}_0}\cos{(\varphi_k-\varphi_b)}}}{\alpha - \beta\cos{(\varphi_k-\varphi^\prime)}} e^{\sigma_b^2\kappa^2 \cos{(\varphi_k-\varphi_p)}} \right)^*
%     \label{int_def2}
% \end{aligned}  
% \end{equation}
% \end{widetext}
%This integral is similar to (\ref{int_def}), so it has the same calculation issues. 
This integral has similar calculation issues to the integral of Eq.~(\ref{int_def}).
Let us see how it behaves in the limit of small $\theta_p$ and $ \beta$ while setting $\vb{b}_0=0$:

\begin{equation}
\begin{aligned}
    \mathcal{I}_{m\sigma\sigma^\prime}^{\sigma_b}(&\alpha,0,0) = \int \frac{\dd\varphi_p}{2\pi}\frac{e^{im\varphi_p}e^{i\sigma(\varphi^\prime-\varphi_p)}}{\alpha} \\
    &\cross \left(\int \frac{\dd\varphi_k}{2\pi}\frac{e^{im\varphi_k}e^{i\sigma^\prime(\varphi^\prime-\varphi_k)}e^{\sigma_b^2\kappa^2 \cos{(\varphi_k-\varphi_p)}}}{\alpha}\right)^* \\
    &= \frac{1}{\alpha^2}\delta_{m-\sigma,m-\sigma^\prime}i^{-(m-\sigma^\prime)}J_{m-\sigma^\prime}(i\sigma_b^2\kappa^2) \\
    &= \frac{1}{\alpha^2}\delta_{\sigma,\sigma^\prime}I_{m-\sigma}(\sigma_b^2\kappa^2),
\end{aligned}
\label{int_sigma_limit}
\end{equation}

\noindent where $I_{m-\sigma}(\sigma_b^2\kappa^2)$ is a modified Bessel function of the first kind. In this limit we find for the amplitude (\ref{amp_meso}):

\begin{equation}
    \begin{aligned}
    \int&\abs{F^{(m)}_{\lambda,\lambda^\prime}(\vb{p},\vb{p}^\prime,\vb{b})}^2 \frac{1}{2\pi\sigma_b^2}e^{-\frac{1}{2}\left(\frac{\vb{b}}{\sigma_b}\right)^2}\dd^2 b =\\
    &= 
    \frac{2Z^2e^4}{\pi} \kappa \left(\varepsilon^2\delta_{\lambda,\lambda^\prime}+m_e^2\delta_{\lambda,-\lambda^\prime}\right) e^{-\sigma_b^2\kappa^2} \\
    &\cross \sum\limits_{\sigma = \pm 1/2} \left(d_{\sigma,\lambda}^{1/2}(\theta_p)d_{\sigma,\lambda^\prime}^{1/2}(\theta^\prime)\right)^2 \frac{1}{\alpha^2}I_{m-\sigma}(\sigma_b^2\kappa^2) \\
    &= \frac{2Z^2e^4}{\pi} \kappa \left(\varepsilon^2\cos{(\theta^\prime/2)}\delta_{\lambda,\lambda^\prime}+m_e^2\sin{(\theta^\prime/2)}\delta_{\lambda,-\lambda^\prime}\right) \\
    &\cross e^{-\sigma_b^2\kappa^2} \frac{1}{\alpha^2}I_{m-\lambda}(\sigma_b^2\kappa^2).
    \end{aligned}
\end{equation}

And for the number of events we have

\begin{equation}
\begin{aligned}
    &\left(\frac{\dd \nu}{\dd \Omega}\right)^{(meso)} = \frac{N_e}{16\pi^2}\frac{\abs{\vb{p}}}{\varepsilon} \abs{F_{m,\lambda,\lambda^\prime}^{(meso)}(\vb{p},\vb{p}^\prime,\vb{b}_0)}^2 \\
    &= 2Z^2e^4\frac{\kappa}{R}  \frac{N_e}{\cos{\theta_p}}\left(\varepsilon^2\delta_{\lambda,\lambda^\prime}+m_e^2\delta_{\lambda,-\lambda^\prime}\right) e^{-\sigma_b^2\kappa^2} \\
    &\sum\limits_{\sigma,\sigma^\prime = \pm 1/2} d_{\sigma,\lambda}^{1/2}(\theta_p)d_{\sigma,\lambda^\prime}^{1/2}(\theta^\prime)d_{\sigma^\prime,\lambda}^{1/2}(\theta_p)d_{\sigma^\prime,\lambda^\prime}^{1/2}(\theta^\prime)\mathcal{I}_{m\sigma\sigma^\prime}^{\sigma_b}(\alpha,\beta,\vb{b}_0),
\end{aligned}
\label{noe_meso_1}
\end{equation}

\bigbreak

\end{appendix}

\bigbreak

\bibliography{main.bib}

\end{document}